\documentclass[aps,preprint,amssymb,superscriptaddress]{revtex4-1}
\usepackage{floatrow}
\newfloatcommand{capbtabbox}{table}[\FBwidth]
\usepackage{graphicx}% Include figure files
\usepackage{dcolumn}% Align table columns on decimal point
\usepackage{bm}% bold math
\usepackage{natbib}
\usepackage{CJK}
\usepackage[utf8]{inputenc}
\usepackage[T1]{fontenc}
\usepackage{mathptmx}
\usepackage{gensymb}
\usepackage{graphicx}% Include figure files
\usepackage{epstopdf}
\usepackage{dcolumn}% Align table columns on decimal point
\usepackage{bm}% bold math
\usepackage{amsmath}
\usepackage{mathtools}
\usepackage{relsize}
\usepackage{multirow}
\usepackage[colorlinks]{hyperref}
\begin{document}

%\preprint{AIP/123-QED}

%\begin{CJK*}{GB}{}
%\preprint{APS/123-QED}
\title{Charge-State Stability of Color Centers in Wide-Bandgap Semiconductors}% Force line breaks with \\

\author{Rodrick Kuate Defo}
\email{rkuatedefo@princeton.edu}
\affiliation{Department of Electrical and Computer Engineering, Princeton University, Princeton, NJ 08540}
\author{Alejandro W. Rodriguez}
\affiliation{Department of Electrical and Computer Engineering, Princeton University, Princeton, NJ 08540}
\author{Steven L. Richardson} 
\affiliation{John A. Paulson School of Engineering and Applied Sciences, Harvard University, Cambridge, MA 02138, USA}
\affiliation{Department of Electrical and Computer Engineering, Howard University, Washington, DC 20059}
\date{\today}% It is always \today, today,

\begin{abstract}
The N$V^-$ color center in diamond has been extensively investigated for quantum sensing, computation, and communication applications. Nonetheless, charge-state decay from the N$V^-$ to its neutral counterpart the N$V^0$ detrimentally affects the robustness of the N$V^-$ center and remains to be fully overcome. In this work, we provide an \textit{ab initio} formalism for accurately estimating the rate of charge-state decay of color centers in wide-bandgap semiconductors. Our formalism employs density functional theory calculations in the context of thermal equilibrium. We illustrate the method using the transition of N$V^-$ to N$V^0$ in the presence of substitutional N [see Z. Yuan \textit{et al}., PRR 2, 033263 (2020)].

%We show in this work that when X$V^-$ color centers exhibit a has been shown to mediate entanglement between electronic spins and surrounding nuclear spins. This entanglement leads to proximal nuclear spins enhancing electronic spin coherence and distant nuclear spins destroying it.   This result suggests that if samples are cooled to a low enough temperature that a Jahn-Teller distortion emerges, if such a distortion does not favor enhanced spin coherence, reheating and repeating the process may ultimately coax the system into a distortion that does. Here the enhancement in spin coherence would be the result of the modulation of the hyperfine interaction strength between the X$V^-$ center electronic spin and the neighboring $^{13}$C nuclei when the center is cycled between the three energetically equivalent distortions. 
\end{abstract}

\maketitle
%\preprint{APS/123-QED}

\section{INTRODUCTION}
Color centers in wide-bandgap semiconductor hosts have garnered significant interest for potential applications in quantum computation~\cite{Childress2013diamond,Weber2010quantum,Pezzagna2021quantum}, communication~\cite{Childress2013diamond,Su2009high,Bradac2019quantum}, and sensing~\cite{Zhang2021toward}. The N$V^-$ color center in diamond in particular, consisting of a single substitutional N atom adjacent to a C vacancy with an additional negative charge, has enjoyed several research advances including the realization of a coherence time on the order of seconds~\cite{Bar-Gill2013solid} and entanglement of N$V^-$ pairs over a distance greater than a kilometer~\cite{Hensen2015loophole}. A drawback in the utilization of N$V^-$ centers for computation, communication, and sensing applications, however, is their tendency to revert to the neutral state after optical initialization into the singly negatively charged state~\cite{Yuan2020charge}. The implied large hole-capture cross section of optically activated N$V^-$ centers in diamond has been investigated both experimentally and theoretically~\cite{Yuan2020charge,Lozovoi2021optical,Lozovoi2023detection}, where the theory has included a bound-exciton model~\cite{Lozovoi2021optical} and semi-classical Monte Carlo simulations~\cite{Lozovoi2023detection}. Arriving at a first-principles description of carrier capture that can apply in thermal equilibrium or under the action of an external field is still the subject of much work~\cite{Chen2023semi,Lozovoi2023detection,Alkauskas2014first,Shi2015comparative,Turiansky2021nonrad,Barmparis2015theory}. Herein we provide such a framework which is accurate and demonstrates that ionizing-dopant concentrations along with the electronic structure of the color center of interest and of the ionizing dopant are crucial for the determination of the expected timescale for hole capture by the ionized color centers. To provide an illustration of the method, we investigate charge transfer rates for N$V^-$ centers~\cite{Degen,Doherty2013the,Rondin_2014,Schirhagl,Kurtsiefer,Jelezko,Gruber2012,Balasubramanian2009ultra,Childress,Hao2023first} in the presence of N in diamond. We note that the approach also applies to Si$V^-$~\cite{neu2011single,Kuate2021calculating,haussler2017photoluminescence,ekimov2019effect}, Ge$V^-$~\cite{Ekimov2015germanium,Kuate2021calculating,Iwasaki2015germanium,Ralchenko2015observation,Ekimov2017anharmonicity,Palyanov2016high,Palyanov2015germanium,Bhaskar2017quantum,bray2018single}, Sn$V^-$~\cite{Iwasaki2017tin,Kuate2021calculating,Ekimov2018tin,tchernij2017single,palyanov2019high, Alkahtani2018tin, rugar2019char, wahl2020direct,fukuta2021sn,Gorlitz2020spectroscopic,Trusheim2020transform,Rugar2021quantum}, and Pb$V^-$\cite{Trusheim2019lead,Kuate2021calculating, tchernij2018single} centers in diamond or color centers in other wide-bandgap semiconductors~\cite{Castelletto2014a,Bockstedte2004ab,Kimoto2014fundamentals,Kuate2018energetics,Gadalla2021enhanced,Kuate2019parallel,Bracher2017selective,Soykal2016silicon,Soykal2017quantum,Weber2010quantum,Gali2011time,Awschalom2018quantum,Wolfowicz2021quantum,Whiteley2019spin}.

This organization of this work is the following. Our computational tools are presented in Section \ref{sec:methodology}. We then present our theoretical approach to charge transfer mediated by built-in electric fields in Section \ref{sec:methods} and provide our results and a discussion of our results for the rate of charge-state decay in Section \ref{sec:imp}. Finally, our conclusions are presented in Section \ref{sec:conc}.

\section{Computational Methods}\label{sec:methodology}
We used VASP~\cite{Kresse1993ab,Kresse1996efficient,Kuate2021theor,Kresse1999from} for our formation energy calculations with the screened HSE06 hybrid functional for exchange and correlation ~\cite{Heyd,Krukau}. We terminated our calculations when the forces in the atomic-position relaxations dropped below a threshold of $10^{-2}$ eV$\cdot$\AA$^{-1}$. The wavefunctions were expanded in a planewave basis with a cutoff energy of 430~eV and the size of the supercell was 512 atoms ($4\times4\times4$ multiple of the conventional unit cell). We employed $\Gamma$-point integration to evaluate energies. The elements used in our calculations and the associated ground-state structures and values of their chemical potentials are: N ($\beta$ hexagonal close-packed structure, $-11.39$ eV/atom) and C (diamond structure, $-11.28$ eV/atom). Only ground-state formation energies were computed in this work.
 
\section{Theoretical Approach} \label{sec:methods} 
\subsection{Thermally driven charge transfer}
As in earlier work~\cite{Kuate2023theor}, we consider the expected rate of transfer for an electron associated with a defect in a crystal, but no longer in the fully dilute limit. Suppose the electron has a definite momentum at each point in time $\hbar\mathbf{k}(t)$. In the non-relativistic limit, we can compute the rate associated with the transfer of the electron with effective mass $m^*$ across a distance $|\Delta\mathbf{r}|$ between ionized defects A and D as 
\begin{align}
\Gamma &=\int_{t_0}^\infty\text{d}t\frac{1}{t-t_0}\frac{\left|\hbar\mathbf{k}(t)\cdot\Delta\mathbf{r}\right|}{m^*|\Delta\mathbf{r}|^2}\int_{t_0}^t\text{d}t^\prime\frac{\hbar\mathbf{k}(t^\prime)\cdot\Delta\mathbf{r}}{m^*|\Delta\mathbf{r}|^2}\\&\times\delta\left(1-\frac{\int_{ t_0}^t\text{d}\tilde{t}\hbar\mathbf{k}(\tilde{t})\cdot\Delta\mathbf{r}}{m^*|\Delta\mathbf{r}|^2}\right).\nonumber
\end{align}
Above, we have used the definition of the Dirac delta function to convey that the rate is obtained by considering the average projection of the velocity onto the desired displacement of the electron divided by the distance $|\Delta\mathbf{r}|$ at the time the electron has completed the trajectory. 

The likelihood that the electron has enough initial kinetic energy to overcome the ionizing potential and recover the neutral system is given by a Fermi-Dirac distribution. We consider the donor level $E_{\text{F}_\text{D}}$ to be associated with an ionized donor-dopant D and the acceptor level $E_{\text{F}_\text{A}}$ to be associated with an ionized color center A. We will show below that the required initial kinetic energy $K_{\text{initial}}$ is equal to the difference between the adiabatic charge-transition levels $E_{\text{F}_\text{D}}$ and $E_{\text{F}_\text{A}}$ for defects D and A respectively, up to a correction of order $k_BT$. We can therefore write the expected rate of charge transfer as
 
\begin{align}
\label{eq:Gammakk}
\left<\Gamma\right> &\approx\int_{t_0}^\infty\text{d}t\frac{1}{t-t_0}\frac{\left|\hbar\mathbf{k}(t)\cdot\Delta\mathbf{r}\right|}{m^*|\Delta\mathbf{r}|^2}\int_{t_0}^t\text{d}t^\prime\frac{\hbar\mathbf{k}(t^\prime)\cdot\Delta\mathbf{r}}{m^*|\Delta\mathbf{r}|^2}\\&\times\frac{\delta\left(1-\frac{\int_{ t_0}^t\text{d}\tilde{t}\hbar\mathbf{k}(\tilde{t})\cdot\Delta\mathbf{r}}{m^*|\Delta\mathbf{r}|^2}\right)}{\exp\left((E_{\text{F}_\text{D}}-E_{\text{F}_\text{A}})/k_BT\right)+1},\nonumber
\end{align}
where $T$ is the temperature and $k_B$ is the Boltzmann constant. The ionization probability is modeled through the Fermi-Dirac distribution since we consider thermal processes for the charge recombination and not processes involving optical excitation.

%and
%\begin{align}
%\label{eq:hGamma}
%    \bar{\Gamma}_h &= \sum_{\text{D}~\in~\text{Donors}}N_{\text{D}}\int_{V_\text{D}}\text{d}\mathbf{r}\rho_\text{D}(\mathbf{r})\sum_{\text{A}~\in~\text{Acceptors}}N_{\text{A}}\int_{V_\text{A}}\text{d}\mathbf{r}^\prime\rho_\text{A}(\mathbf{r}^\prime)\frac{1}{N_s}\sum_{s}\Omega_{PUC}\int\frac{\text{d}\mathbf{k}(t_0)}{(2\pi)^3}\int\frac{\text{d}\mathbf{k}^\prime}{(2\pi)^3}\\&\int_{t_0}^\infty\text{d}t\frac{1}{t-t_0}\frac{1}{\hbar}\frac{|\nabla_\mathbf{k}\epsilon^\text{V}_{\mathbf{k}(t),s}\cdot(\mathbf{r}-\mathbf{r}^\prime)|}{|\mathbf{r}-\mathbf{r}^\prime|^2}\int_{t_0}^t\text{d}t^\prime\frac{1}{\hbar}\frac{\nabla_\mathbf{k}\epsilon^\text{V}_{\mathbf{k}(t^\prime),s}\cdot(\mathbf{r}-\mathbf{r}^\prime)}{|\mathbf{r}-\mathbf{r}^\prime|^2}\nonumber\\&\times\frac{\delta\left(1-\frac{\int_{t_0}^t\text{d}\tilde{t}\frac{1}{\hbar}\nabla_\mathbf{k}\epsilon^\text{V}_{\mathbf{k}(\tilde{t}),s}\cdot(\mathbf{r}-\mathbf{r}^\prime)}{|\mathbf{r}-\mathbf{r}^\prime|^2}\right)}{\exp((\epsilon^\text{V}_{\mathbf{k}^\prime,s}-E_{\text{F}_\text{D}})/k_BT)+1}\nonumber\times\frac{\delta(\mathbf{k}^\prime-\mathbf{k}(t_0)-\frac{1}{\hbar}\int_{t_0}^t\mathbf{F}_h(\tilde{t}^\prime)\text{d}\tilde{t}^\prime)}{\exp((E_{\text{F}_\text{A}}-\epsilon^\text{V}_{\mathbf{k}(t_0),s})/k_BT)+1}\nonumber.
%\end{align}

\subsection{Introducing electric forces}
The calculation of the built-in electric field between defect pairs, which counteracts the electromagnetic potential that results in their ionization, requires the evaluation of electrostatic potentials. In order to compute the total energies of the systems from which the electrostatic potentials will be obtained, we solve for the eigenvalues of the Hamiltonian~\cite{Kaxiras2003atomic}
\begin{align}
\mathcal{H} &= -\sum_I\frac{\hbar^2}{2M_I}\nabla^2_{\mathbf{R}_I} - \sum_i\frac{\hbar^2}{2m_e}\nabla^2_{\mathbf{r}_i}\\&-\sum_{iI}\frac{Z_Ie^2}{|\mathbf{R}_I-\mathbf{r}_i|}+\frac{1}{2}\sum_{ij(j\neq i)}\frac{e^2}{|\mathbf{r}_i-\mathbf{r}_j|}+\frac{1}{2}\sum_{IJ(J\neq I)}\frac{Z_IZ_Je^2}{|\mathbf{R}_I-\mathbf{R}_J|}\nonumber
\end{align}
where $\mathbf{r}_i$ and $m_e$ denote the position and rest mass of electron $i$, respectively, and $\mathbf{R}_I$, $Z_I$, and $M_I$ are the position, valence charge, and rest mass of ion $I$, respectively. We apply the Born-Oppenheimer approximation to decouple the electronic and ionic degrees of freedom and solve for the electronic degrees of freedom given static ionic positions. If the eigenvalues for the electronic Hamiltonian have zero dispersion in reciprocal space, the expectation value of the velocity of the electrons will be zero implying that we can neglect the kinetic terms. Therefore, for defects introduced into a crystal in the dilute limit where the dispersion of the donor and acceptor levels is vanishingly small, our Hamiltonian effectively captures Coulomb or electrostatic potentials. 

The formation energies $\Delta H_f({\rm X^{\rm q}}, \, \{\mu_i^\text{X}\}, \, E_\text{F})$~\cite{zhang1991chemical, Freysoldt2014first,Kuate2018energetics,Kuate2019how,Kuate2021methods,Kuate2021theor,zunger2021under,Yang2015self,Ashcroft1976solid,Kuate2023theor} calculated in the dilute limit for defects capture electrostatic potentials as a result of the aforementioned arguments and would therefore allow for the determination of built-in fields. In $\Delta H_f({\rm X^{\rm q}}, \, \{\mu_i^\text{X}\}, \, E_\text{F})$, X is the defect species for which the formation energy is being calculated, q is the charge state of X, $\{\mu_i^\text{X}\}$ is the set of chemical potentials for the constituents of X, and $E_\text{F}$ is the Fermi level. In order to consistently determine the potentials at the locations of the defect species and to be in line with standard conventions, we use the neutral system in each case as the reference for the zero of the energy. The potential due to each charged defect consequently becomes the difference between the energies of the charged-defect containing system and the neutral system divided by the compensating background charge. The potential associated with a defect X with charge q is then simply
\begin{equation}
\label{eq:phiX}
    \phi(\mathbf{r}_\text{X}) = -\frac{1}{e\text{q}}\left(\Delta H_f({\rm X^{\rm q}}, \, \{\mu_i^\text{X}\}, \, 0)-\Delta H_f({\rm X^{0}}, \, \{\mu_i^\text{X}\}, \, 0)\right)
\end{equation}
and similarly for a defect Y with charge $-$q the potential follows from (\ref{eq:phiX}) upon the substitution $\text{X} \to \text{Y}$ and $\text{q} \to -\text{q}$,
%\begin{equation}
%\label{eq:phiY}
%    \phi(\mathbf{r}_\text{Y}) = \frac{1}{e\text{q}}\left(\Delta H_f({\rm Y^{-\rm q}}, \, \{\mu_i^\text{Y}\}, \, 0)-\Delta H_f({\rm Y^{0}}, \, \{\mu_i^\text{Y}\}, \, 0)\right)
%\end{equation}
with $\mathbf{r}_\text{X}$ and $\mathbf{r}_\text{Y}$ denoting the locations of the respective defects. Here, $E_\text{F}$ is set to zero since its inclusion in the expression for the formation energy subtracts out the energy associated with adding charge to the defect, which is no longer necessary if we are employing the electrostatic energy corresponding to a given charged defect. Above we have also treated the defects in the dilute limit so that the compensating background charge of the computational supercell can be treated as a point charge relative to the entire crystal. The built-in field at the location of the defect Y is then given by
\begin{align}
\label{eq:Efield}
    \vec{\mathcal{E}} = -\nabla(\phi(\mathbf{r})) \approx -\frac{(\phi(\mathbf{r}_\text{Y})-\phi(\mathbf{r}_\text{X}))}{|\Delta\mathbf{r}|}\frac{\Delta\mathbf{r}}{|\Delta\mathbf{r}|},
\end{align}
where $\Delta\mathbf{r} = \mathbf{r}_\text{Y}-\mathbf{r}_\text{X}$. 

We demonstrate the equivalence with our earlier work~\cite{Kuate2023theor} as follows. In our earlier work~\cite{Kuate2023theor}, we provided the expression
\begin{equation}
    \label{eq:Efieldequiv}
    \vec{\mathcal{E}} \approx \frac{1}{e}\frac{E_\text{V}(\mathbf{r}_\text{A})-E_\text{V}(\mathbf{r}_\text{D}))}{|\Delta\mathbf{r}|}\frac{\Delta\mathbf{r}}{|\Delta\mathbf{r}|},
\end{equation}
where $\mathbf{r}_\text{A}$ denoted the location of an acceptor A, $\mathbf{r}_\text{D}$ denoted the location of a donor D, $\Delta\mathbf{r} = \mathbf{r}_\text{A}-\mathbf{r}_\text{D}$, and $E_\text{V}$ indicated the energy of the valence band extremum. We had found that
\begin{align}
    E_\text{V}(\mathbf{r}_\text{A})&-E_\text{V}(\mathbf{r}_\text{D}) =\\&\bigg(\Delta H_f({\rm D^{0}}, \, \{\mu_i^\text{D}\}, \, 0)-\Delta H_f({\rm D^{+}}, \, \{\mu_i^\text{D}\}, \, 0)\bigg)\nonumber\\-&\bigg(\Delta H_f({\rm A^{-}}, \, \{\mu_i^\text{A}\}, \, 0)-\Delta H_f({\rm A^{0}}, \, \{\mu_i^\text{A}\}, \, 0)\bigg).\nonumber
\end{align}
Setting $\text{q} = 1$ implicitly in Eq. (\ref{eq:Efield}), we find agreement between Eq. (\ref{eq:Efield}) and Eq. (\ref{eq:Efieldequiv}) demonstrating the desired equivalence between the `local' Fermi level and built-in electric-field formulations.
%formulation and this formulation in terms of the built-in electric field between two point charges. 

\subsection{Computing the momentum due to electric forces}
At every point in time the charge must have enough energy to sustain motion along the path between the defects. In order to compute the necessary momentum for an electron moving between the defects, we recognize that the momentum must be sufficient to overcome the ionizing electromagnetic potential energy. We note, however, that as the charge approaches defect Y from defect X a built-in field will emerge to cancel the ionizing potential. In order to capture the emergence of the built-in field, we employ an energy conservation argument. Effectively, our argument is that the sum of the potential energy leading to the ionization of the defect pair and the kinetic energy of the electron must be conserved, so that $U + K = U(K=0)$. Thus, one finds
\begin{align}
%U + K &= U(K = 0) \\
U &= U(K=0) - K \\
&= e\left(\phi(\mathbf{r}_\text{X})-\phi(\mathbf{r}_\text{Y})\right) - \frac{\hbar^{2} k^2}{2m^*}.\label{eq:Udef}
\end{align}
Above, the electromagnetic potential energy, $U(K=0)$, reflects the relative formation energies associated with the placement of the electron that will travel from defect X to defect Y. A more general potential would allow for the inclusion of arbitrary external fields. The value of the potential energy landscape is defined in the manner given in Eq. (\ref{eq:Udef}) at the location of defect X so that its value at the location of defect Y can be set to zero.

In order to explicitly determine the required momentum as a function of time, we apply the second law of motion and employ a discrete approximation given the small distances with the origin at the location of defect Y to obtain
\begin{align}
\frac{d(\hbar \mathbf{k})}{dt} = -\nabla U \approx -\frac{U}{|\mathbf{r}|}\frac{\mathbf{r}}{|\mathbf{r}|}.
\end{align}
The non-relativistic limit has been applied above, which is justified by the fact that the maximum speed an electron can attain according to our calculations is less than $0.002c$, where $c$ is the speed of light. Once the electron arrives at defect Y from defect X, the kinetic energy would be dissipated in a process akin to the M\"ossbauer effect so that there is no need to produce a large initial change in momentum. Therefore, given an initial speed corresponding to a kinetic energy $E_{\text{thermal}} = k_BT$ in a random direction at an angle $\theta$ with respect to the radial direction
\begin{align}
\label{eq:drdt}
\frac{dr}{dt} \approx \pm \sqrt{2/m^*\frac{\left(k_BT\cos^2(\theta)-U(K=0)\ln(r/r_0)\right)}{1-\ln(r/r_0)}}.
\end{align}
After writing $\hbar \mathbf{k} = m^*\frac{\text{d}\mathbf{r}}{\text{d}t}$, the relation $\frac{d^2r}{dt^2} = \frac{1}{2}\frac{d\left(\frac{dr}{dt}\right)^2}{dr}$ has been used to obtain Eq. (\ref{eq:drdt}). We neglect $\frac{d\theta}{dt}$ and $\frac{d^2\theta}{dt^2}$ since in our work $k_BT \ll U(K=0) = e\left(\phi(\mathbf{r}_\text{X})-\phi(\mathbf{r}_\text{Y})\right) = E_{\text{F}_{\text{D}}} - E_{\text{F}_{\text{A}}}$, where we have used the definition of the donor and acceptor levels~\cite{Kuate2023theor}. 

In previous work~\cite{Kuate2023theor} we considered transport via extended states, resulting in capture cross sections that would have finite extent if the wavevectors of the charges were allowed to evolve under the influence of the corresponding electron or hole forces. Without wavevector evolution, the capture cross sections would be points. The wavevector evolution employed in this work for charges in extended states under the influence of forces is consistent with the experimental finding of a large hole-capture cross section for optically activated N$V^-$ centers~\cite{Yuan2020charge,Lozovoi2021optical,Lozovoi2023detection}. Inclusion of the $\frac{d\theta}{dt}$ and $\frac{d^2\theta}{dt^2}$ terms would allow for further refinement of the capture cross sections.

Therefore, if the electron is located at defect Y with an initial kinetic energy $E_{\text{F}_{\text{D}}} - E_{\text{F}_{\text{A}}}$ up to a correction of order $k_BT$, then the electron will be able to return to defect X. For a nonzero gradient $U(K=0) > 0$ and zero initial velocity, subsequent $r$ will be less than $r_0 = |\Delta \mathbf{r}|$, requiring a velocity given by applying the negative root in Eq. (\ref{eq:drdt}). We note that the required kinetic energy can equivalently be viewed as the energy needed to excite the excitonic system from its ground state corresponding to the ionized defect pair where the charges are bound to the defects to its excited stated corresponding to the defect pair where the charges are free from the defects and the defects are neutral, in which case the momentum simply follows from the electric force between the charges.

We can average the reciprocal of Eq. (\ref{eq:Gammakk}) over all possible initial velocity directions, given an initial thermal energy of $E_{\text{thermal}} = k_BT$, which yields 
\begin{align}
\label{eq:tildeGammakk}
    \left<\tau\right> &\approx \frac{1}{\pi}\int_0^\pi\text{d}\theta\Biggl(\frac{1}{\Delta t}\\&\times\frac{1}{\left(\exp\left(\left(E_{\text{F}_\text{D}}-E_{\text{F}_\text{A}}-E_\text{initial}\right)/k_BT\right)+1\right)}\Biggr)^{-1}.\nonumber
\end{align}
Above, $\Delta t = \int_{r_0}^0\text{d}r\left(\frac{dr}{dt}\right)^{-1}$ and $E_\text{initial} = k_BT\cos^2(\theta)\cdot\text{sgn}(\cos(\theta))$. We account for phonon excitations by the introduction of the thermal correction $E_\text{initial}$. The introduction of the correction of order $k_BT$ influences the barrier for charge transfer in a manner analogous to the effect of phonons on the absorption energy of a fluorescent defect, namely the averaged timescale $\left<\tau\right>$ behaves as if the defect pair were experiencing a larger barrier for charge transfer than without the thermal correction.

In more detail, the barrier at zero temperature (in the absence of phonons) is given by $E_{\text{F}_{\text{D}}} - E_{\text{F}_{\text{A}}}$. At finite temperature (in the presence of phonons), the barrier is shifted by  $-E_\text{initial} = -k_BT\cos^2(\theta)\cdot\text{sgn}(\cos(\theta))$. Averaging this shift over all $\theta$ (from $\theta = 0$ to $\theta = \pi$) should give zero for the net shift of the barrier. The barrier is in the argument of an exponential, however, so that the contributions when $\text{sgn}(\cos(\theta))$ is negative carry more weight in the average than the contributions when $\text{sgn}(\cos(\theta))$ is positive. Thus, the effective barrier (the barrier we would obtain if we replaced the timescale expression that averages over $\theta$ with a timescale expression that employs a single fixed value for the barrier) is higher than the barrier obtained by neglecting the thermal correction. Indeed, at a population fraction of approximately 50\% for N$V^-$ centers, the upper and lower estimates for the charge-state decay time computed without the thermal correction yield values that are approximately 3.98 times smaller than the values reported in Fig. \ref{fig:comparison}. Therefore, the room temperature (300~K) contribution of phonons is to increase the barrier by approximately $k_BT\ln(3.98) \approx 0.036$~eV.

Care must be applied if $\theta > \pi/2$ in which case we must first integrate with positive velocity from $r = r_0$ to $r = r_0\exp(k_BT\cdot \cos^2(\theta)/U(K=0))$ and then back from $r = r_0\exp(k_BT\cdot \cos^2(\theta)/U(K=0))$ to $r = r_0$ with negative velocity before performing the integral between $r= r_0$ and $r= 0$ with negative velocity. 

In order to obtain the fraction of color centers that have undergone charge-state decay for a given timescale $\left<\tau\right>$, we compute the probability that a color center will have undergone charge-decay as
\begin{align}
\label{eq:prob}
P(\tau) &= \frac{8}{l_\text{X}^3}\bigg|B_{r_0}[\mathbf{0}]\\&\cap\left\{\mathbf{x}\in\mathbb{R}^3 : \max_{i=1,2}\left\{\left|x_i-\frac{l^{3/2}_\text{X}}{\sqrt{32}d^{1/2}_{\text{max}}}\right|\right\}\leq \frac{l^{3/2}_\text{X}}{\sqrt{32}d^{1/2}_{\text{max}}}\right\}\nonumber\\&\cap\left\{\mathbf{x}\in\mathbb{R}^3 : |x_3-d_{\text{max}}/2|\leq d_{\text{max}}/2\right\}\bigg|\nonumber
\end{align}
Above, $|\cdot|$ denotes the volume of the enclosed region, $B_{r_0}[\mathbf{0}]$ is the closed ball of radius $r_0$ centered at the origin,  $l_{\text{X}} = n_\text{X}^{-1/3}$ where $n_\text{X}$ is the concentration of the ionizing dopant X, and $d_\text{max}$ is the maximum implantation depth, since we consider the case where $d_\text{max} < l_{\text{X}}/2$ in this work. The probability reflects the fact that since $\left<\tau\right>$ is monotonic in $r_0$ the fraction of color centers having undergone charge state decay for a given $\left<\tau\right>$ corresponding to a given $r_0$ will be given by the fraction of color centers that are separated from their ionizing dopant by a distance of $r_0$ or less. We evaluate Eqs. (\ref{eq:tildeGammakk}) and (\ref{eq:prob}) for $r_0$ uniformly distributed between $r_0 = a$ and $r_0 = \sqrt{\frac{l_\text{X}^3}{4d_\text{max}}+d_\text{max}^2}$, where $a$ is the lattice constant of the conventional unit cell of diamond ($a = 3.549$~\AA~\cite{Kuate2023theor}). 

\section{Results and Discussion}\label{sec:imp}
\subsection{Experimentally investigating charge-state decay of N$V^-$}
The details of the experiment of Yuan \textit{et al.}~\cite{Yuan2020charge} investigating charge-state instability of near-surface N$V^-$ centers in  diamond are as follows. A green pulse was used to initialize N$V$ centers in the negative charge state. The N$V^-$ centers were then left under darkness for a variable delay time following which they were read out using a charge-state-selective orange pulse. The samples used included a diamond sample labeled A that was implanted with an implantation dose of $5\times10^8$~cm$^{-2}$ at an energy of 3~keV and that was polished, pre-etched, and $^{12}$C enriched and a diamond sample labeled F that was implanted with an implantation dose of $1\times10^9$~cm$^{-2}$ at an energy of 1.5~keV and that was polished and pre-etched. They found that charge-state decay for sample F occurred on a timescale from 11-300~ms, while for sample A much less decay was observed out to 1~s. The accelerated charge conversion for sample F was attributed to the availability of electron traps near the surface, in particular boron impurities. In the following, we therefore concern ourselves with the theoretical determination of the charge-state decay rate in sample A. 

\subsection{Effect of defect species and concentration on charge-conversion timescales}
\begin{figure}[ht!] 
\centering
\includegraphics[width=0.99\textwidth]{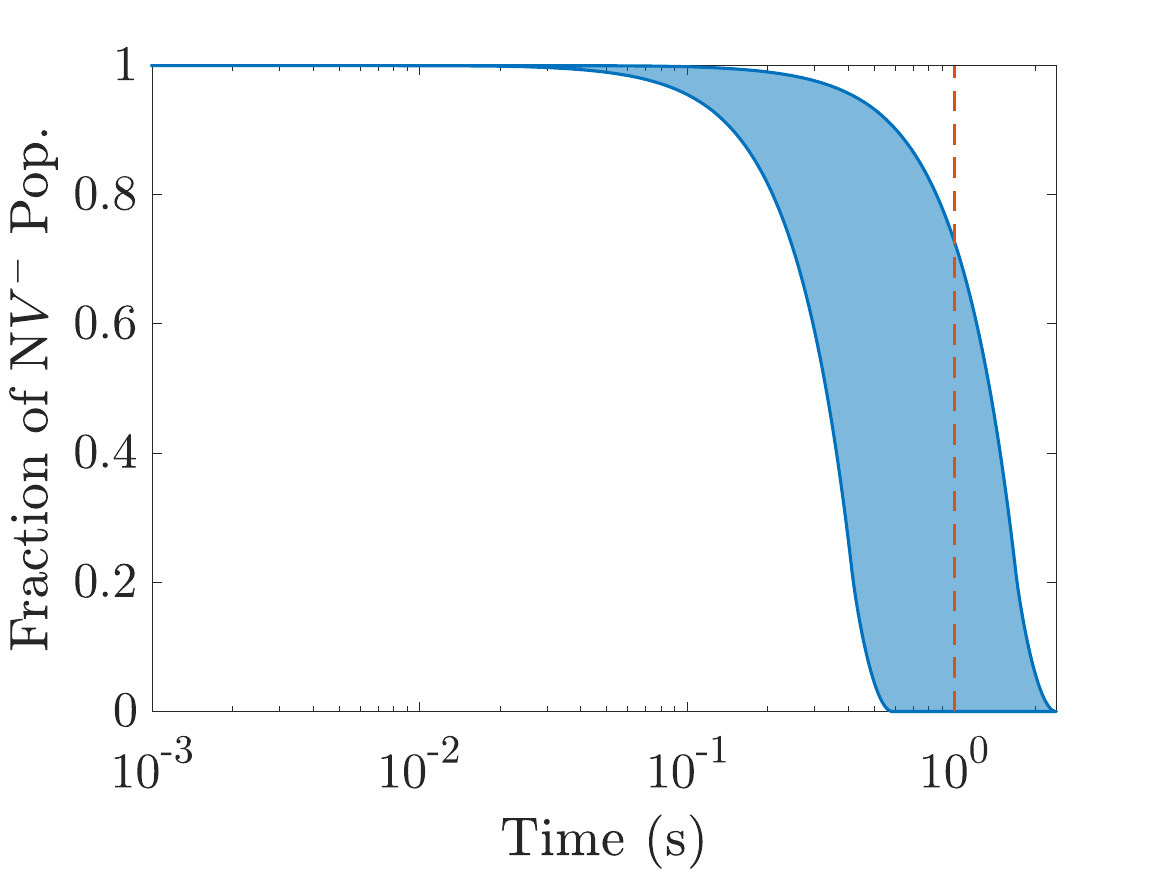}
\caption{Fraction of N$V^-$ population remaining as a function of time, calculated via Eqs. (\ref{eq:tildeGammakk}) and (\ref{eq:prob}) with $T = 300$~K. We model charge as undergoing transfer from N$V^-$ to N$_\text{C}^+$. The ionizing dopant concentration is $n_\text{X} = \frac{5\times10^8~\text{cm}^{-2}}{d_{\text{max}}}$, using an estimated maximum implantation depth of $d_\text{max} = 3.5~\text{nm}\cdot\text{keV}^{-1} E_\text{imp}$ where $E_\text{imp}$ is the implantation energy~\cite{Broadway2018spat}. We employ $E_\text{imp} = 3$~keV~\cite{Yuan2020charge}. The light-blue shaded region represents the result of introducing a shift in $E_{\text{F}_\text{A}}$ of $0.03~\text{eV} \lesssim \Delta E_{\text{F}_\text{A}} \lesssim 0.07~\text{eV}$ and of applying the transformation $E_{\text{F}_\text{D}} \rightarrow (E_{\text{F}_\text{D}} + 3.4~\text{eV})/2$. The orange dashed line indicates experimental results for the timescale associated with charge-state decay of four representative individual N$V^-$ centers in sample A with final relative populations between $p = 0.76$ and $p = 0.93$~\cite{Yuan2020charge}. Since the four individual N$V^-$ centers were representative, the fraction of the N$V^-$ population remaining after the charge-state decay of each center can lie anywhere between 0 and 1.} 
\label{fig:comparison}
\end{figure}

In order to determine timescales for charge transfer between ionized defect species, we apply Eqs. (\ref{eq:tildeGammakk}) and (\ref{eq:prob}) for an ionized color center with acceptor level $E_{\text{F}_\text{A}}$ transferring an electron to an ionized donor dopant with donor level $E_{\text{F}_\text{D}}$. The acceptor and donor levels serving as $E_{\text{F}_\text{A}}$ and $E_{\text{F}_\text{D}}$, measured relative to the valence band maximum, are given by $\epsilon^{\text{N}V}(0/-)\approx 2.8$~eV and $\epsilon^{\text{N}_\text{C}}(0/+)\approx 3.6$~eV, respectively. An effective mass of $m^*\approx 1.48m_e$ is obtained from fitting the bandstructure in Ref.~\cite{Kuate2023theor}. We account for the fact that the charge-state decay of  individual N$V^-$ centers measured in the Yuan~\textit{et al.} experiment was to a steady-state relative population greater than 0.5~\cite{Yuan2020charge}, where 0.5 is the value corresponding to local pinning of the Fermi level at $\epsilon^{\text{N}V}(0/-)$, by introducing a shift in $E_{\text{F}_\text{A}}$ such that the local Fermi level $E_{\text{F}_\text{A}}$ would produce a relative N$V^-$ population that would correspond to the experimentally measured value. Explicitly, for a final relative population of $p$, the shift is $\Delta E_{\text{F}_\text{A}} = k_BT\ln\left(\frac{(1-p_0)p}{p_0(1-p)}\right)$, where $p_0 = 0.5$. This result is obtained by considering the relative populations of the charge states of a single N$V$ defect. Since we are considering the relative populations for a single defect, we can drop the contribution from configurational entropy. Therefore, the relative population of the N$V^-$ state is given by
\begin{equation}
p = \frac{\exp(-(\Delta H_f(\text{N}V^-,\{\mu_i^{\text{N}V}\},0)-E_\text{F})/k_BT)}{\exp(-(\Delta H_f(\text{N}V^-,\{\mu_i^{\text{N}V}\},0)-E_\text{F})/k_BT)+\exp(-\Delta H_f(\text{N}V^0,\{\mu_i^{\text{N}V}\},0)/k_BT)},
\end{equation}
where we have assumed that the relative populations of the charge states other than $0$ and $-1$ are negligible since the Fermi level is pinned near the $\epsilon^{\text{N}V}(0/-)$ charge-transition level and have used the relation $\Delta H_f(\text{N}V^-,\{\mu_i^{\text{N}V}\},E_\text{F}) = \Delta H_f(\text{N}V^-,\{\mu_i^{\text{N}V}\},0) - E_\text{F}$~\cite{Kuate2023theor}. If we define
\begin{equation}
p_0 = \frac{\exp(-(\Delta H_f(\text{N}V^-,\{\mu_i^{\text{N}V}\},0)-\epsilon^{\text{N}V}(0/-))/k_BT}{\exp(-(\Delta H_f(\text{N}V^-,\{\mu_i^{\text{N}V}\},0)-\epsilon^{\text{N}V}(0/-))/k_BT)+\exp(-\Delta H_f(\text{N}V^0,\{\mu_i^{\text{N}V}\},0)/k_BT)},
\end{equation}
and write $E_\text{F} =  \Delta E_\text{F} + \epsilon^{\text{N}V}(0/-)$, then it follows immediately that 
\begin{equation}
\Delta E_{\text{F}} = k_BT\ln\left(\frac{(1-p_0)p}{p_0(1-p)}\right).
\end{equation}
The maximum shift is $\Delta E_{\text{F}_\text{A}} \approx 0.07$~eV and the minimum shift is $\Delta E_{\text{F}_\text{A}} \approx 0.03$~eV. The corresponding values of $p$ are $0.93$ and $0.76$, respectively~\cite{Yuan2020charge}. We also account for the effect of the surface, which we assume to have an ether-like termination~\cite{Sque2006structure}, so that a donor level measured relative to the valence band maximum of 3.4~eV~\cite{Sque2006structure,Broadway2018spat} is induced. For N$V^-$ centers near the surface the number of ionized surface donors should be commensurate with the number of ionized bulk donors~\cite{Broadway2018spat}, so we can employ an effective $E_{\text{F}_\text{D}}$ given by averaging the donor-level values ($E_{\text{F}_\text{D}} \rightarrow (E_{\text{F}_\text{D}} + 3.4~\text{eV})/2)$. Such averaging would also bring our earlier results in better agreement with experimental results at shallow implantation depths~\cite{Kuate2023theor,Broadway2018spat}. The result of these corrections produces good agreement with experiment (see Fig. \ref{fig:comparison}).

Explicitly, the averaging of the bulk N$_\text{C}$ defect level, $E_{\text{F}_\text{D}}$, and the ether-like surface termination defect level of 3.4~eV follows from assuming that on average only one N$_\text{C}$ defect and one ether-like surface defect will equilibrate after each excitation of the sample needed to perform a measurement and before the N$V$ itself has time to equilibrate with the dopant from which it obtained its charge. The assumption is justified by that fact that in the model proposed in this work, which considers the limit that is not fully dilute (meaning that charge does not need to fully enter the band edges to travel between defects), as well as in the model from Ref.~\cite{Kuate2023theor}, where charge must enter the band edges to travel between defects, the Fermi-level equilibration timescale will be the approximately the same for the N$V$-N$_\text{C}$ defect pair as for the ether-like defect-N$_\text{C}$ defect pair. This result follows from the fact that in the case of Ref.~\cite{Kuate2023theor} the energy that governs the rate of equilibration is the difference between the energy of the conduction band minimum and the energy of the N$_\text{C}$ defect level (the defect levels of the ether-like surface defect and the N$V$ can be neglected since they are both lower in energy than the N$_\text{C}$ defect level and the defects therefore act as acceptors for which the Femi-Dirac distribution is approximately equal to 1). In the case of this work, since the energy of the N$_\text{C}$ defect level is higher than that of either the N$V$ or the ether-like surface defect, the factor of the Fermi-Dirac distribution yields approximately 1 for the ionization process and the speeds at which the charge travels between the N$_\text{C}$ and the N$V$ and between the N$_\text{C}$ and the ether-like surface defect differ by a factor on the order of unity since the speeds evolve as the square root of energies that differ by a factor of order unity. By contrast, if the 
measurement timescale is such that more than one ether-like surface defect is allowed to equilibrate with more than one N$_\text{C}$ defect, then the value of the Fermi level is obtained from more general charge conservation considerations (Eq. (36) in Ref.~\cite{Freysoldt2014first}). 

\clearpage
\section{Conclusions} \label{sec:conc}
We have shown that the precise electronic structure of an ionizing-dopant species and of an ionized color center are highly relevant to the charge-state decay characteristics of the ionized color center. The concentrations of dopants and color centers are also integral to the elucidation of charge-transfer rates within a semiconductor sample.  A key implication of our results is that, in order to mitigate charge-state decay for ionized color centers in semiconductors, the color center should be chosen such that its charge-transition level lies much lower in energy than the donor level of the ionizing dopant.  

R.K.D. gratefully acknowledges financial support from the Princeton Presidential Postdoctoral Research Fellowship and from the National Academies of Science, Engineering, and Medicine Ford Foundation Postdoctoral Fellowship program. We additionally acknowledge support by the STC Center for Integrated Quantum Materials, NSF Grant No. DMR-1231319. Finally, we wish to acknowledge insightful feedback from Nathalie P. de Leon and fruitful discussions with Pengning Chao. We also thank the referees for their many critical and helpful suggestions which have been instrumental in improving the clarity of our paper.

\bibliography{refs_NV}

\begin{thebibliography}{81}%
\makeatletter
\providecommand \@ifxundefined [1]{%
 \@ifx{#1\undefined}
}%
\providecommand \@ifnum [1]{%
 \ifnum #1\expandafter \@firstoftwo
 \else \expandafter \@secondoftwo
 \fi
}%
\providecommand \@ifx [1]{%
 \ifx #1\expandafter \@firstoftwo
 \else \expandafter \@secondoftwo
 \fi
}%
\providecommand \natexlab [1]{#1}%
\providecommand \enquote  [1]{``#1''}%
\providecommand \bibnamefont  [1]{#1}%
\providecommand \bibfnamefont [1]{#1}%
\providecommand \citenamefont [1]{#1}%
\providecommand \href@noop [0]{\@secondoftwo}%
\providecommand \href [0]{\begingroup \@sanitize@url \@href}%
\providecommand \@href[1]{\@@startlink{#1}\@@href}%
\providecommand \@@href[1]{\endgroup#1\@@endlink}%
\providecommand \@sanitize@url [0]{\catcode `\\12\catcode `\$12\catcode
  `\&12\catcode `\#12\catcode `\^12\catcode `\_12\catcode `\%12\relax}%
\providecommand \@@startlink[1]{}%
\providecommand \@@endlink[0]{}%
\providecommand \url  [0]{\begingroup\@sanitize@url \@url }%
\providecommand \@url [1]{\endgroup\@href {#1}{\urlprefix }}%
\providecommand \urlprefix  [0]{URL }%
\providecommand \Eprint [0]{\href }%
\providecommand \doibase [0]{https://doi.org/}%
\providecommand \selectlanguage [0]{\@gobble}%
\providecommand \bibinfo  [0]{\@secondoftwo}%
\providecommand \bibfield  [0]{\@secondoftwo}%
\providecommand \translation [1]{[#1]}%
\providecommand \BibitemOpen [0]{}%
\providecommand \bibitemStop [0]{}%
\providecommand \bibitemNoStop [0]{.\EOS\space}%
\providecommand \EOS [0]{\spacefactor3000\relax}%
\providecommand \BibitemShut  [1]{\csname bibitem#1\endcsname}%
\let\auto@bib@innerbib\@empty
%</preamble>
\bibitem [{\citenamefont {Childress}\ and\ \citenamefont
  {Hanson}(2013)}]{Childress2013diamond}%
  \BibitemOpen
  \bibfield  {author} {\bibinfo {author} {\bibfnamefont {L.}~\bibnamefont
  {Childress}}\ and\ \bibinfo {author} {\bibfnamefont {R.}~\bibnamefont
  {Hanson}},\ }\href {https://doi.org/10.1557/mrs.2013.20} {\bibfield
  {journal} {\bibinfo  {journal} {MRS Bulletin}\ }\textbf {\bibinfo {volume}
  {38}},\ \bibinfo {pages} {134} (\bibinfo {year} {2013})}\BibitemShut
  {NoStop}%
\bibitem [{\citenamefont {Weber}\ \emph {et~al.}(2010)\citenamefont {Weber},
  \citenamefont {Koehl}, \citenamefont {Varley}, \citenamefont {Janotti},
  \citenamefont {Buckley}, \citenamefont {de~Walle},\ and\ \citenamefont
  {Awschalom}}]{Weber2010quantum}%
  \BibitemOpen
  \bibfield  {author} {\bibinfo {author} {\bibfnamefont {J.~R.}\ \bibnamefont
  {Weber}}, \bibinfo {author} {\bibfnamefont {W.~F.}\ \bibnamefont {Koehl}},
  \bibinfo {author} {\bibfnamefont {J.~B.}\ \bibnamefont {Varley}}, \bibinfo
  {author} {\bibfnamefont {A.}~\bibnamefont {Janotti}}, \bibinfo {author}
  {\bibfnamefont {B.~B.}\ \bibnamefont {Buckley}}, \bibinfo {author}
  {\bibfnamefont {C.~G.~V.}\ \bibnamefont {de~Walle}},\ and\ \bibinfo {author}
  {\bibfnamefont {D.~D.}\ \bibnamefont {Awschalom}},\ }\href
  {https://doi.org/10.1073/pnas.1003052107} {\bibfield  {journal} {\bibinfo
  {journal} {Proceedings of the National Academy of Sciences}\ }\textbf
  {\bibinfo {volume} {107}},\ \bibinfo {pages} {8513} (\bibinfo {year}
  {2010})}\BibitemShut {NoStop}%
\bibitem [{\citenamefont {Pezzagna}\ and\ \citenamefont
  {Meijer}(2021)}]{Pezzagna2021quantum}%
  \BibitemOpen
  \bibfield  {author} {\bibinfo {author} {\bibfnamefont {S.}~\bibnamefont
  {Pezzagna}}\ and\ \bibinfo {author} {\bibfnamefont {J.}~\bibnamefont
  {Meijer}},\ }\href {https://doi.org/10.1063/5.0007444} {\bibfield  {journal}
  {\bibinfo  {journal} {Applied Physics Reviews}\ }\textbf {\bibinfo {volume}
  {8}},\ \bibinfo {pages} {011308} (\bibinfo {year} {2021})}\BibitemShut
  {NoStop}%
\bibitem [{\citenamefont {Su}\ \emph {et~al.}(2009)\citenamefont {Su},
  \citenamefont {Greentree},\ and\ \citenamefont {Hollenberg}}]{Su2009high}%
  \BibitemOpen
  \bibfield  {author} {\bibinfo {author} {\bibfnamefont {C.-H.}\ \bibnamefont
  {Su}}, \bibinfo {author} {\bibfnamefont {A.~D.}\ \bibnamefont {Greentree}},\
  and\ \bibinfo {author} {\bibfnamefont {L.~C.~L.}\ \bibnamefont
  {Hollenberg}},\ }\href {https://doi.org/10.1103/PhysRevA.80.052308}
  {\bibfield  {journal} {\bibinfo  {journal} {Phys. Rev. A}\ }\textbf {\bibinfo
  {volume} {80}},\ \bibinfo {pages} {052308} (\bibinfo {year}
  {2009})}\BibitemShut {NoStop}%
\bibitem [{\citenamefont {Bradac}\ \emph {et~al.}(2019)\citenamefont {Bradac},
  \citenamefont {Gao}, \citenamefont {Forneris}, \citenamefont {Trusheim},\
  and\ \citenamefont {Aharonovich}}]{Bradac2019quantum}%
  \BibitemOpen
  \bibfield  {author} {\bibinfo {author} {\bibfnamefont {C.}~\bibnamefont
  {Bradac}}, \bibinfo {author} {\bibfnamefont {W.}~\bibnamefont {Gao}},
  \bibinfo {author} {\bibfnamefont {J.}~\bibnamefont {Forneris}}, \bibinfo
  {author} {\bibfnamefont {M.~E.}\ \bibnamefont {Trusheim}},\ and\ \bibinfo
  {author} {\bibfnamefont {I.}~\bibnamefont {Aharonovich}},\ }\href
  {https://doi.org/10.1038/s41467-019-13332-w} {\bibfield  {journal} {\bibinfo
  {journal} {Nature Communications}\ }\textbf {\bibinfo {volume} {10}},\
  \bibinfo {pages} {5625} (\bibinfo {year} {2019})}\BibitemShut {NoStop}%
\bibitem [{\citenamefont {Zhang}\ \emph {et~al.}(2021)\citenamefont {Zhang},
  \citenamefont {Pramanik}, \citenamefont {Zhang}, \citenamefont {Gulka},
  \citenamefont {Wang}, \citenamefont {Jing}, \citenamefont {Xu}, \citenamefont
  {Li}, \citenamefont {Wei}, \citenamefont {Cigler},\ and\ \citenamefont
  {Chu}}]{Zhang2021toward}%
  \BibitemOpen
  \bibfield  {author} {\bibinfo {author} {\bibfnamefont {T.}~\bibnamefont
  {Zhang}}, \bibinfo {author} {\bibfnamefont {G.}~\bibnamefont {Pramanik}},
  \bibinfo {author} {\bibfnamefont {K.}~\bibnamefont {Zhang}}, \bibinfo
  {author} {\bibfnamefont {M.}~\bibnamefont {Gulka}}, \bibinfo {author}
  {\bibfnamefont {L.}~\bibnamefont {Wang}}, \bibinfo {author} {\bibfnamefont
  {J.}~\bibnamefont {Jing}}, \bibinfo {author} {\bibfnamefont {F.}~\bibnamefont
  {Xu}}, \bibinfo {author} {\bibfnamefont {Z.}~\bibnamefont {Li}}, \bibinfo
  {author} {\bibfnamefont {Q.}~\bibnamefont {Wei}}, \bibinfo {author}
  {\bibfnamefont {P.}~\bibnamefont {Cigler}},\ and\ \bibinfo {author}
  {\bibfnamefont {Z.}~\bibnamefont {Chu}},\ }\href
  {https://doi.org/10.1021/acssensors.1c00415} {\bibfield  {journal} {\bibinfo
  {journal} {ACS Sensors}\ }\textbf {\bibinfo {volume} {6}},\ \bibinfo {pages}
  {2077} (\bibinfo {year} {2021})}\BibitemShut {NoStop}%
\bibitem [{\citenamefont {Bar-Gill}\ \emph {et~al.}(2013)\citenamefont
  {Bar-Gill}, \citenamefont {Pham}, \citenamefont {Jarmola}, \citenamefont
  {Budker},\ and\ \citenamefont {Walsworth}}]{Bar-Gill2013solid}%
  \BibitemOpen
  \bibfield  {author} {\bibinfo {author} {\bibfnamefont {N.}~\bibnamefont
  {Bar-Gill}}, \bibinfo {author} {\bibfnamefont {L.~M.}\ \bibnamefont {Pham}},
  \bibinfo {author} {\bibfnamefont {A.}~\bibnamefont {Jarmola}}, \bibinfo
  {author} {\bibfnamefont {D.}~\bibnamefont {Budker}},\ and\ \bibinfo {author}
  {\bibfnamefont {R.~L.}\ \bibnamefont {Walsworth}},\ }\href
  {https://doi.org/10.1038/ncomms2771} {\bibfield  {journal} {\bibinfo
  {journal} {Nature Communications}\ }\textbf {\bibinfo {volume} {4}},\
  \bibinfo {pages} {1743} (\bibinfo {year} {2013})}\BibitemShut {NoStop}%
\bibitem [{\citenamefont {Hensen}\ \emph {et~al.}(2015)\citenamefont {Hensen},
  \citenamefont {Bernien}, \citenamefont {Dr{\'e}au}, \citenamefont {Reiserer},
  \citenamefont {Kalb}, \citenamefont {Blok}, \citenamefont {Ruitenberg},
  \citenamefont {Vermeulen}, \citenamefont {Schouten}, \citenamefont
  {Abell{\'a}n}, \citenamefont {Amaya}, \citenamefont {Pruneri}, \citenamefont
  {Mitchell}, \citenamefont {Markham}, \citenamefont {Twitchen}, \citenamefont
  {Elkouss}, \citenamefont {Wehner}, \citenamefont {Taminiau},\ and\
  \citenamefont {Hanson}}]{Hensen2015loophole}%
  \BibitemOpen
  \bibfield  {author} {\bibinfo {author} {\bibfnamefont {B.}~\bibnamefont
  {Hensen}}, \bibinfo {author} {\bibfnamefont {H.}~\bibnamefont {Bernien}},
  \bibinfo {author} {\bibfnamefont {A.~E.}\ \bibnamefont {Dr{\'e}au}}, \bibinfo
  {author} {\bibfnamefont {A.}~\bibnamefont {Reiserer}}, \bibinfo {author}
  {\bibfnamefont {N.}~\bibnamefont {Kalb}}, \bibinfo {author} {\bibfnamefont
  {M.~S.}\ \bibnamefont {Blok}}, \bibinfo {author} {\bibfnamefont
  {J.}~\bibnamefont {Ruitenberg}}, \bibinfo {author} {\bibfnamefont {R.~F.~L.}\
  \bibnamefont {Vermeulen}}, \bibinfo {author} {\bibfnamefont {R.~N.}\
  \bibnamefont {Schouten}}, \bibinfo {author} {\bibfnamefont {C.}~\bibnamefont
  {Abell{\'a}n}}, \bibinfo {author} {\bibfnamefont {W.}~\bibnamefont {Amaya}},
  \bibinfo {author} {\bibfnamefont {V.}~\bibnamefont {Pruneri}}, \bibinfo
  {author} {\bibfnamefont {M.~W.}\ \bibnamefont {Mitchell}}, \bibinfo {author}
  {\bibfnamefont {M.}~\bibnamefont {Markham}}, \bibinfo {author} {\bibfnamefont
  {D.~J.}\ \bibnamefont {Twitchen}}, \bibinfo {author} {\bibfnamefont
  {D.}~\bibnamefont {Elkouss}}, \bibinfo {author} {\bibfnamefont
  {S.}~\bibnamefont {Wehner}}, \bibinfo {author} {\bibfnamefont {T.~H.}\
  \bibnamefont {Taminiau}},\ and\ \bibinfo {author} {\bibfnamefont
  {R.}~\bibnamefont {Hanson}},\ }\href {https://doi.org/10.1038/nature15759}
  {\bibfield  {journal} {\bibinfo  {journal} {Nature}\ }\textbf {\bibinfo
  {volume} {526}},\ \bibinfo {pages} {682} (\bibinfo {year}
  {2015})}\BibitemShut {NoStop}%
\bibitem [{\citenamefont {Yuan}\ \emph {et~al.}(2020)\citenamefont {Yuan},
  \citenamefont {Fitzpatrick}, \citenamefont {Rodgers}, \citenamefont
  {Sangtawesin}, \citenamefont {Srinivasan},\ and\ \citenamefont
  {de~Leon}}]{Yuan2020charge}%
  \BibitemOpen
  \bibfield  {author} {\bibinfo {author} {\bibfnamefont {Z.}~\bibnamefont
  {Yuan}}, \bibinfo {author} {\bibfnamefont {M.}~\bibnamefont {Fitzpatrick}},
  \bibinfo {author} {\bibfnamefont {L.~V.~H.}\ \bibnamefont {Rodgers}},
  \bibinfo {author} {\bibfnamefont {S.}~\bibnamefont {Sangtawesin}}, \bibinfo
  {author} {\bibfnamefont {S.}~\bibnamefont {Srinivasan}},\ and\ \bibinfo
  {author} {\bibfnamefont {N.~P.}\ \bibnamefont {de~Leon}},\ }\href
  {https://doi.org/10.1103/PhysRevResearch.2.033263} {\bibfield  {journal}
  {\bibinfo  {journal} {Phys. Rev. Research}\ }\textbf {\bibinfo {volume}
  {2}},\ \bibinfo {pages} {033263} (\bibinfo {year} {2020})}\BibitemShut
  {NoStop}%
\bibitem [{\citenamefont {Lozovoi}\ \emph {et~al.}(2021)\citenamefont
  {Lozovoi}, \citenamefont {Jayakumar}, \citenamefont {Daw}, \citenamefont
  {Vizkelethy}, \citenamefont {Bielejec}, \citenamefont {Doherty},
  \citenamefont {Flick},\ and\ \citenamefont {Meriles}}]{Lozovoi2021optical}%
  \BibitemOpen
  \bibfield  {author} {\bibinfo {author} {\bibfnamefont {A.}~\bibnamefont
  {Lozovoi}}, \bibinfo {author} {\bibfnamefont {H.}~\bibnamefont {Jayakumar}},
  \bibinfo {author} {\bibfnamefont {D.}~\bibnamefont {Daw}}, \bibinfo {author}
  {\bibfnamefont {G.}~\bibnamefont {Vizkelethy}}, \bibinfo {author}
  {\bibfnamefont {E.}~\bibnamefont {Bielejec}}, \bibinfo {author}
  {\bibfnamefont {M.~W.}\ \bibnamefont {Doherty}}, \bibinfo {author}
  {\bibfnamefont {J.}~\bibnamefont {Flick}},\ and\ \bibinfo {author}
  {\bibfnamefont {C.~A.}\ \bibnamefont {Meriles}},\ }\href
  {https://doi.org/10.1038/s41928-021-00656-z} {\bibfield  {journal} {\bibinfo
  {journal} {Nature Electronics}\ }\textbf {\bibinfo {volume} {4}},\ \bibinfo
  {pages} {717} (\bibinfo {year} {2021})}\BibitemShut {NoStop}%
\bibitem [{\citenamefont {Lozovoi}\ \emph {et~al.}(2023)\citenamefont
  {Lozovoi}, \citenamefont {Chen}, \citenamefont {Vizkelethy}, \citenamefont
  {Bielejec}, \citenamefont {Flick}, \citenamefont {Doherty},\ and\
  \citenamefont {Meriles}}]{Lozovoi2023detection}%
  \BibitemOpen
  \bibfield  {author} {\bibinfo {author} {\bibfnamefont {A.}~\bibnamefont
  {Lozovoi}}, \bibinfo {author} {\bibfnamefont {Y.}~\bibnamefont {Chen}},
  \bibinfo {author} {\bibfnamefont {G.}~\bibnamefont {Vizkelethy}}, \bibinfo
  {author} {\bibfnamefont {E.}~\bibnamefont {Bielejec}}, \bibinfo {author}
  {\bibfnamefont {J.}~\bibnamefont {Flick}}, \bibinfo {author} {\bibfnamefont
  {M.~W.}\ \bibnamefont {Doherty}},\ and\ \bibinfo {author} {\bibfnamefont
  {C.~A.}\ \bibnamefont {Meriles}},\ }\href
  {https://doi.org/10.1021/acs.nanolett.3c00860} {\bibfield  {journal}
  {\bibinfo  {journal} {Nano Letters}\ }\textbf {\bibinfo {volume} {23}},\
  \bibinfo {pages} {4495} (\bibinfo {year} {2023})}\BibitemShut {NoStop}%
\bibitem [{\citenamefont {Chen}\ \emph {et~al.}(2023)\citenamefont {Chen},
  \citenamefont {Oberg}, \citenamefont {Flick}, \citenamefont {Lozovoi},
  \citenamefont {Meriles},\ and\ \citenamefont {Doherty}}]{Chen2023semi}%
  \BibitemOpen
  \bibfield  {author} {\bibinfo {author} {\bibfnamefont {Y.}~\bibnamefont
  {Chen}}, \bibinfo {author} {\bibfnamefont {L.}~\bibnamefont {Oberg}},
  \bibinfo {author} {\bibfnamefont {J.}~\bibnamefont {Flick}}, \bibinfo
  {author} {\bibfnamefont {A.}~\bibnamefont {Lozovoi}}, \bibinfo {author}
  {\bibfnamefont {C.~A.}\ \bibnamefont {Meriles}},\ and\ \bibinfo {author}
  {\bibfnamefont {M.~W.}\ \bibnamefont {Doherty}},\ }\href@noop {} {\bibinfo
  {title} {Semi-$\textit{ab initio}$ modeling of bound states of deep defects
  in semiconductors}} (\bibinfo {year} {2023}),\ \Eprint
  {https://arxiv.org/abs/2306.12005} {arXiv:2306.12005 [cond-mat.mes-hall]}
  \BibitemShut {NoStop}%
\bibitem [{\citenamefont {Alkauskas}\ \emph {et~al.}(2014)\citenamefont
  {Alkauskas}, \citenamefont {Yan},\ and\ \citenamefont {Van~de
  Walle}}]{Alkauskas2014first}%
  \BibitemOpen
  \bibfield  {author} {\bibinfo {author} {\bibfnamefont {A.}~\bibnamefont
  {Alkauskas}}, \bibinfo {author} {\bibfnamefont {Q.}~\bibnamefont {Yan}},\
  and\ \bibinfo {author} {\bibfnamefont {C.~G.}\ \bibnamefont {Van~de Walle}},\
  }\href {https://doi.org/10.1103/PhysRevB.90.075202} {\bibfield  {journal}
  {\bibinfo  {journal} {Phys. Rev. B}\ }\textbf {\bibinfo {volume} {90}},\
  \bibinfo {pages} {075202} (\bibinfo {year} {2014})}\BibitemShut {NoStop}%
\bibitem [{\citenamefont {Shi}\ \emph {et~al.}(2015)\citenamefont {Shi},
  \citenamefont {Xu},\ and\ \citenamefont {Wang}}]{Shi2015comparative}%
  \BibitemOpen
  \bibfield  {author} {\bibinfo {author} {\bibfnamefont {L.}~\bibnamefont
  {Shi}}, \bibinfo {author} {\bibfnamefont {K.}~\bibnamefont {Xu}},\ and\
  \bibinfo {author} {\bibfnamefont {L.-W.}\ \bibnamefont {Wang}},\ }\href
  {https://doi.org/10.1103/PhysRevB.91.205315} {\bibfield  {journal} {\bibinfo
  {journal} {Phys. Rev. B}\ }\textbf {\bibinfo {volume} {91}},\ \bibinfo
  {pages} {205315} (\bibinfo {year} {2015})}\BibitemShut {NoStop}%
\bibitem [{\citenamefont {Turiansky}\ \emph {et~al.}(2021)\citenamefont
  {Turiansky}, \citenamefont {Alkauskas}, \citenamefont {Engel}, \citenamefont
  {Kresse}, \citenamefont {Wickramaratne}, \citenamefont {Shen}, \citenamefont
  {Dreyer},\ and\ \citenamefont {{Van de Walle}}}]{Turiansky2021nonrad}%
  \BibitemOpen
  \bibfield  {author} {\bibinfo {author} {\bibfnamefont {M.~E.}\ \bibnamefont
  {Turiansky}}, \bibinfo {author} {\bibfnamefont {A.}~\bibnamefont
  {Alkauskas}}, \bibinfo {author} {\bibfnamefont {M.}~\bibnamefont {Engel}},
  \bibinfo {author} {\bibfnamefont {G.}~\bibnamefont {Kresse}}, \bibinfo
  {author} {\bibfnamefont {D.}~\bibnamefont {Wickramaratne}}, \bibinfo {author}
  {\bibfnamefont {J.-X.}\ \bibnamefont {Shen}}, \bibinfo {author}
  {\bibfnamefont {C.~E.}\ \bibnamefont {Dreyer}},\ and\ \bibinfo {author}
  {\bibfnamefont {C.~G.}\ \bibnamefont {{Van de Walle}}},\ }\href
  {https://doi.org/https://doi.org/10.1016/j.cpc.2021.108056} {\bibfield
  {journal} {\bibinfo  {journal} {Computer Physics Communications}\ }\textbf
  {\bibinfo {volume} {267}},\ \bibinfo {pages} {108056} (\bibinfo {year}
  {2021})}\BibitemShut {NoStop}%
\bibitem [{\citenamefont {Barmparis}\ \emph {et~al.}(2015)\citenamefont
  {Barmparis}, \citenamefont {Puzyrev}, \citenamefont {Zhang},\ and\
  \citenamefont {Pantelides}}]{Barmparis2015theory}%
  \BibitemOpen
  \bibfield  {author} {\bibinfo {author} {\bibfnamefont {G.~D.}\ \bibnamefont
  {Barmparis}}, \bibinfo {author} {\bibfnamefont {Y.~S.}\ \bibnamefont
  {Puzyrev}}, \bibinfo {author} {\bibfnamefont {X.-G.}\ \bibnamefont {Zhang}},\
  and\ \bibinfo {author} {\bibfnamefont {S.~T.}\ \bibnamefont {Pantelides}},\
  }\href {https://doi.org/10.1103/PhysRevB.92.214111} {\bibfield  {journal}
  {\bibinfo  {journal} {Phys. Rev. B}\ }\textbf {\bibinfo {volume} {92}},\
  \bibinfo {pages} {214111} (\bibinfo {year} {2015})}\BibitemShut {NoStop}%
\bibitem [{\citenamefont {Degen}(2008)}]{Degen}%
  \BibitemOpen
  \bibfield  {author} {\bibinfo {author} {\bibfnamefont {C.~L.}\ \bibnamefont
  {Degen}},\ }\href {https://doi.org/10.1063/1.2943282} {\bibfield  {journal}
  {\bibinfo  {journal} {Applied Physics Letters}\ }\textbf {\bibinfo {volume}
  {92}},\ \bibinfo {pages} {243111} (\bibinfo {year} {2008})}\BibitemShut
  {NoStop}%
\bibitem [{\citenamefont {Doherty}\ \emph {et~al.}(2013)\citenamefont
  {Doherty}, \citenamefont {Manson}, \citenamefont {Delaney}, \citenamefont
  {Jelezko}, \citenamefont {Wrachtrup},\ and\ \citenamefont
  {Hollenberg}}]{Doherty2013the}%
  \BibitemOpen
  \bibfield  {author} {\bibinfo {author} {\bibfnamefont {M.~W.}\ \bibnamefont
  {Doherty}}, \bibinfo {author} {\bibfnamefont {N.~B.}\ \bibnamefont {Manson}},
  \bibinfo {author} {\bibfnamefont {P.}~\bibnamefont {Delaney}}, \bibinfo
  {author} {\bibfnamefont {F.}~\bibnamefont {Jelezko}}, \bibinfo {author}
  {\bibfnamefont {J.}~\bibnamefont {Wrachtrup}},\ and\ \bibinfo {author}
  {\bibfnamefont {L.~C.}\ \bibnamefont {Hollenberg}},\ }\href
  {https://doi.org/https://doi.org/10.1016/j.physrep.2013.02.001} {\bibfield
  {journal} {\bibinfo  {journal} {Physics Reports}\ }\textbf {\bibinfo {volume}
  {528}},\ \bibinfo {pages} {1 } (\bibinfo {year} {2013})}\BibitemShut
  {NoStop}%
\bibitem [{\citenamefont {Rondin}\ \emph {et~al.}(2014)\citenamefont {Rondin},
  \citenamefont {Tetienne}, \citenamefont {Hingant}, \citenamefont {Roch},
  \citenamefont {Maletinsky},\ and\ \citenamefont {Jacques}}]{Rondin_2014}%
  \BibitemOpen
  \bibfield  {author} {\bibinfo {author} {\bibfnamefont {L.}~\bibnamefont
  {Rondin}}, \bibinfo {author} {\bibfnamefont {J.-P.}\ \bibnamefont
  {Tetienne}}, \bibinfo {author} {\bibfnamefont {T.}~\bibnamefont {Hingant}},
  \bibinfo {author} {\bibfnamefont {J.-F.}\ \bibnamefont {Roch}}, \bibinfo
  {author} {\bibfnamefont {P.}~\bibnamefont {Maletinsky}},\ and\ \bibinfo
  {author} {\bibfnamefont {V.}~\bibnamefont {Jacques}},\ }\href
  {https://doi.org/10.1088/0034-4885/77/5/056503} {\bibfield  {journal}
  {\bibinfo  {journal} {Reports on Progress in Physics}\ }\textbf {\bibinfo
  {volume} {77}},\ \bibinfo {pages} {056503} (\bibinfo {year}
  {2014})}\BibitemShut {NoStop}%
\bibitem [{\citenamefont {Schirhagl}\ \emph {et~al.}(2014)\citenamefont
  {Schirhagl}, \citenamefont {Chang}, \citenamefont {Loretz},\ and\
  \citenamefont {Degen}}]{Schirhagl}%
  \BibitemOpen
  \bibfield  {author} {\bibinfo {author} {\bibfnamefont {R.}~\bibnamefont
  {Schirhagl}}, \bibinfo {author} {\bibfnamefont {K.}~\bibnamefont {Chang}},
  \bibinfo {author} {\bibfnamefont {M.}~\bibnamefont {Loretz}},\ and\ \bibinfo
  {author} {\bibfnamefont {C.~L.}\ \bibnamefont {Degen}},\ }\href
  {https://doi.org/10.1146/annurev-physchem-040513-103659} {\bibfield
  {journal} {\bibinfo  {journal} {Annual Review of Physical Chemistry}\
  }\textbf {\bibinfo {volume} {65}},\ \bibinfo {pages} {83} (\bibinfo {year}
  {2014})}\BibitemShut {NoStop}%
\bibitem [{\citenamefont {Kurtsiefer}\ \emph {et~al.}(2000)\citenamefont
  {Kurtsiefer}, \citenamefont {Mayer}, \citenamefont {Zarda},\ and\
  \citenamefont {Weinfurter}}]{Kurtsiefer}%
  \BibitemOpen
  \bibfield  {author} {\bibinfo {author} {\bibfnamefont {C.}~\bibnamefont
  {Kurtsiefer}}, \bibinfo {author} {\bibfnamefont {S.}~\bibnamefont {Mayer}},
  \bibinfo {author} {\bibfnamefont {P.}~\bibnamefont {Zarda}},\ and\ \bibinfo
  {author} {\bibfnamefont {H.}~\bibnamefont {Weinfurter}},\ }\href
  {https://doi.org/10.1103/PhysRevLett.85.290} {\bibfield  {journal} {\bibinfo
  {journal} {Phys. Rev. Lett.}\ }\textbf {\bibinfo {volume} {85}},\ \bibinfo
  {pages} {290} (\bibinfo {year} {2000})}\BibitemShut {NoStop}%
\bibitem [{\citenamefont {Jelezko}\ \emph {et~al.}(2004)\citenamefont
  {Jelezko}, \citenamefont {Gaebel}, \citenamefont {Popa}, \citenamefont
  {Domhan}, \citenamefont {Gruber},\ and\ \citenamefont {Wrachtrup}}]{Jelezko}%
  \BibitemOpen
  \bibfield  {author} {\bibinfo {author} {\bibfnamefont {F.}~\bibnamefont
  {Jelezko}}, \bibinfo {author} {\bibfnamefont {T.}~\bibnamefont {Gaebel}},
  \bibinfo {author} {\bibfnamefont {I.}~\bibnamefont {Popa}}, \bibinfo {author}
  {\bibfnamefont {M.}~\bibnamefont {Domhan}}, \bibinfo {author} {\bibfnamefont
  {A.}~\bibnamefont {Gruber}},\ and\ \bibinfo {author} {\bibfnamefont
  {J.}~\bibnamefont {Wrachtrup}},\ }\href
  {https://doi.org/10.1103/PhysRevLett.93.130501} {\bibfield  {journal}
  {\bibinfo  {journal} {Phys. Rev. Lett.}\ }\textbf {\bibinfo {volume} {93}},\
  \bibinfo {pages} {130501} (\bibinfo {year} {2004})}\BibitemShut {NoStop}%
\bibitem [{\citenamefont {Gruber}\ \emph {et~al.}(1997)\citenamefont {Gruber},
  \citenamefont {Dr{\"a}benstedt}, \citenamefont {Tietz}, \citenamefont
  {Fleury}, \citenamefont {Wrachtrup},\ and\ \citenamefont {von
  Borczyskowski}}]{Gruber2012}%
  \BibitemOpen
  \bibfield  {author} {\bibinfo {author} {\bibfnamefont {A.}~\bibnamefont
  {Gruber}}, \bibinfo {author} {\bibfnamefont {A.}~\bibnamefont
  {Dr{\"a}benstedt}}, \bibinfo {author} {\bibfnamefont {C.}~\bibnamefont
  {Tietz}}, \bibinfo {author} {\bibfnamefont {L.}~\bibnamefont {Fleury}},
  \bibinfo {author} {\bibfnamefont {J.}~\bibnamefont {Wrachtrup}},\ and\
  \bibinfo {author} {\bibfnamefont {C.}~\bibnamefont {von Borczyskowski}},\
  }\href {https://doi.org/10.1126/science.276.5321.2012} {\bibfield  {journal}
  {\bibinfo  {journal} {Science}\ }\textbf {\bibinfo {volume} {276}},\ \bibinfo
  {pages} {2012} (\bibinfo {year} {1997})}\BibitemShut {NoStop}%
\bibitem [{\citenamefont {Balasubramanian}\ \emph {et~al.}(2009)\citenamefont
  {Balasubramanian}, \citenamefont {Neumann}, \citenamefont {Twitchen},
  \citenamefont {Markham}, \citenamefont {Kolesov}, \citenamefont {Mizuochi},
  \citenamefont {Isoya}, \citenamefont {Achard}, \citenamefont {Beck},
  \citenamefont {Tissler}, \citenamefont {Jacques}, \citenamefont {Hemmer},
  \citenamefont {Jelezko},\ and\ \citenamefont
  {Wrachtrup}}]{Balasubramanian2009ultra}%
  \BibitemOpen
  \bibfield  {author} {\bibinfo {author} {\bibfnamefont {G.}~\bibnamefont
  {Balasubramanian}}, \bibinfo {author} {\bibfnamefont {P.}~\bibnamefont
  {Neumann}}, \bibinfo {author} {\bibfnamefont {D.}~\bibnamefont {Twitchen}},
  \bibinfo {author} {\bibfnamefont {M.}~\bibnamefont {Markham}}, \bibinfo
  {author} {\bibfnamefont {R.}~\bibnamefont {Kolesov}}, \bibinfo {author}
  {\bibfnamefont {N.}~\bibnamefont {Mizuochi}}, \bibinfo {author}
  {\bibfnamefont {J.}~\bibnamefont {Isoya}}, \bibinfo {author} {\bibfnamefont
  {J.}~\bibnamefont {Achard}}, \bibinfo {author} {\bibfnamefont
  {J.}~\bibnamefont {Beck}}, \bibinfo {author} {\bibfnamefont {J.}~\bibnamefont
  {Tissler}}, \bibinfo {author} {\bibfnamefont {V.}~\bibnamefont {Jacques}},
  \bibinfo {author} {\bibfnamefont {P.~R.}\ \bibnamefont {Hemmer}}, \bibinfo
  {author} {\bibfnamefont {F.}~\bibnamefont {Jelezko}},\ and\ \bibinfo {author}
  {\bibfnamefont {J.}~\bibnamefont {Wrachtrup}},\ }\href
  {https://doi.org/10.1038/nmat2420} {\bibfield  {journal} {\bibinfo  {journal}
  {Nat. Mater.}\ }\textbf {\bibinfo {volume} {8}},\ \bibinfo {pages} {383}
  (\bibinfo {year} {2009})}\BibitemShut {NoStop}%
\bibitem [{\citenamefont {Childress}\ \emph {et~al.}(2006)\citenamefont
  {Childress}, \citenamefont {Gurudev~Dutt}, \citenamefont {Taylor},
  \citenamefont {Zibrov}, \citenamefont {Jelezko}, \citenamefont {Wrachtrup},
  \citenamefont {Hemmer},\ and\ \citenamefont {Lukin}}]{Childress}%
  \BibitemOpen
  \bibfield  {author} {\bibinfo {author} {\bibfnamefont {L.}~\bibnamefont
  {Childress}}, \bibinfo {author} {\bibfnamefont {M.~V.}\ \bibnamefont
  {Gurudev~Dutt}}, \bibinfo {author} {\bibfnamefont {J.~M.}\ \bibnamefont
  {Taylor}}, \bibinfo {author} {\bibfnamefont {A.~S.}\ \bibnamefont {Zibrov}},
  \bibinfo {author} {\bibfnamefont {F.}~\bibnamefont {Jelezko}}, \bibinfo
  {author} {\bibfnamefont {J.}~\bibnamefont {Wrachtrup}}, \bibinfo {author}
  {\bibfnamefont {P.~R.}\ \bibnamefont {Hemmer}},\ and\ \bibinfo {author}
  {\bibfnamefont {M.~D.}\ \bibnamefont {Lukin}},\ }\href
  {https://doi.org/10.1126/science.1131871} {\bibfield  {journal} {\bibinfo
  {journal} {Science}\ }\textbf {\bibinfo {volume} {314}},\ \bibinfo {pages}
  {281} (\bibinfo {year} {2006})}\BibitemShut {NoStop}%
\bibitem [{\citenamefont {Tang}\ \emph {et~al.}(2023)\citenamefont {Tang},
  \citenamefont {Barr}, \citenamefont {Wang}, \citenamefont {Cappellaro},\ and\
  \citenamefont {Li}}]{Hao2023first}%
  \BibitemOpen
  \bibfield  {author} {\bibinfo {author} {\bibfnamefont {H.}~\bibnamefont
  {Tang}}, \bibinfo {author} {\bibfnamefont {A.~R.}\ \bibnamefont {Barr}},
  \bibinfo {author} {\bibfnamefont {G.}~\bibnamefont {Wang}}, \bibinfo {author}
  {\bibfnamefont {P.}~\bibnamefont {Cappellaro}},\ and\ \bibinfo {author}
  {\bibfnamefont {J.}~\bibnamefont {Li}},\ }\href
  {https://doi.org/10.1021/acs.jpclett.3c00314} {\bibfield  {journal} {\bibinfo
   {journal} {The Journal of Physical Chemistry Letters}\ }\textbf {\bibinfo
  {volume} {14}},\ \bibinfo {pages} {3266} (\bibinfo {year}
  {2023})}\BibitemShut {NoStop}%
\bibitem [{\citenamefont {Neu}\ \emph {et~al.}(2011)\citenamefont {Neu},
  \citenamefont {Steinmetz}, \citenamefont {Riedrich-M\"{o}ller}, \citenamefont
  {Gsell}, \citenamefont {Fischer}, \citenamefont {Schreck},\ and\
  \citenamefont {Becher}}]{neu2011single}%
  \BibitemOpen
  \bibfield  {author} {\bibinfo {author} {\bibfnamefont {E.}~\bibnamefont
  {Neu}}, \bibinfo {author} {\bibfnamefont {D.}~\bibnamefont {Steinmetz}},
  \bibinfo {author} {\bibfnamefont {J.}~\bibnamefont {Riedrich-M\"{o}ller}},
  \bibinfo {author} {\bibfnamefont {S.}~\bibnamefont {Gsell}}, \bibinfo
  {author} {\bibfnamefont {M.}~\bibnamefont {Fischer}}, \bibinfo {author}
  {\bibfnamefont {M.}~\bibnamefont {Schreck}},\ and\ \bibinfo {author}
  {\bibfnamefont {C.}~\bibnamefont {Becher}},\ }\href
  {https://doi.org/10.1088/1367-2630/13/2/025012} {\bibfield  {journal}
  {\bibinfo  {journal} {New J. Phys.}\ }\textbf {\bibinfo {volume} {13}},\
  \bibinfo {pages} {025012} (\bibinfo {year} {2011})}\BibitemShut {NoStop}%
\bibitem [{\citenamefont {Kuate~Defo}\ \emph
  {et~al.}(2021{\natexlab{a}})\citenamefont {Kuate~Defo}, \citenamefont
  {Kaxiras},\ and\ \citenamefont {Richardson}}]{Kuate2021calculating}%
  \BibitemOpen
  \bibfield  {author} {\bibinfo {author} {\bibfnamefont {R.}~\bibnamefont
  {Kuate~Defo}}, \bibinfo {author} {\bibfnamefont {E.}~\bibnamefont
  {Kaxiras}},\ and\ \bibinfo {author} {\bibfnamefont {S.~L.}\ \bibnamefont
  {Richardson}},\ }\href {https://doi.org/10.1103/PhysRevB.104.075158}
  {\bibfield  {journal} {\bibinfo  {journal} {Phys. Rev. B}\ }\textbf {\bibinfo
  {volume} {104}},\ \bibinfo {pages} {075158} (\bibinfo {year}
  {2021}{\natexlab{a}})}\BibitemShut {NoStop}%
\bibitem [{\citenamefont {H{\"a}u{\ss}ler}\ \emph {et~al.}(2017)\citenamefont
  {H{\"a}u{\ss}ler}, \citenamefont {Thiering}, \citenamefont {Dietrich},
  \citenamefont {Waasem}, \citenamefont {Teraji}, \citenamefont {Isoya},
  \citenamefont {Iwasaki}, \citenamefont {Hatano}, \citenamefont {Jelezko},
  \citenamefont {Gali},\ and\ \citenamefont
  {Kubanek}}]{haussler2017photoluminescence}%
  \BibitemOpen
  \bibfield  {author} {\bibinfo {author} {\bibfnamefont {S.}~\bibnamefont
  {H{\"a}u{\ss}ler}}, \bibinfo {author} {\bibfnamefont {G.}~\bibnamefont
  {Thiering}}, \bibinfo {author} {\bibfnamefont {A.}~\bibnamefont {Dietrich}},
  \bibinfo {author} {\bibfnamefont {N.}~\bibnamefont {Waasem}}, \bibinfo
  {author} {\bibfnamefont {T.}~\bibnamefont {Teraji}}, \bibinfo {author}
  {\bibfnamefont {J.}~\bibnamefont {Isoya}}, \bibinfo {author} {\bibfnamefont
  {T.}~\bibnamefont {Iwasaki}}, \bibinfo {author} {\bibfnamefont
  {M.}~\bibnamefont {Hatano}}, \bibinfo {author} {\bibfnamefont
  {F.}~\bibnamefont {Jelezko}}, \bibinfo {author} {\bibfnamefont
  {A.}~\bibnamefont {Gali}},\ and\ \bibinfo {author} {\bibfnamefont
  {A.}~\bibnamefont {Kubanek}},\ }\href
  {https://doi.org/10.1088\%2F1367-2630\%2Faa73e5} {\bibfield  {journal}
  {\bibinfo  {journal} {New J. Phys.}\ }\textbf {\bibinfo {volume} {19}},\
  \bibinfo {pages} {063036} (\bibinfo {year} {2017})}\BibitemShut {NoStop}%
\bibitem [{\citenamefont {Ekimov}\ \emph {et~al.}(2019)\citenamefont {Ekimov},
  \citenamefont {Kondrin}, \citenamefont {Krivobok}, \citenamefont {Khomich},
  \citenamefont {Vlasov}, \citenamefont {Khmelnitskiy}, \citenamefont
  {Iwasaki},\ and\ \citenamefont {Hatano}}]{ekimov2019effect}%
  \BibitemOpen
  \bibfield  {author} {\bibinfo {author} {\bibfnamefont {E.}~\bibnamefont
  {Ekimov}}, \bibinfo {author} {\bibfnamefont {M.}~\bibnamefont {Kondrin}},
  \bibinfo {author} {\bibfnamefont {V.}~\bibnamefont {Krivobok}}, \bibinfo
  {author} {\bibfnamefont {A.}~\bibnamefont {Khomich}}, \bibinfo {author}
  {\bibfnamefont {I.}~\bibnamefont {Vlasov}}, \bibinfo {author} {\bibfnamefont
  {R.}~\bibnamefont {Khmelnitskiy}}, \bibinfo {author} {\bibfnamefont
  {T.}~\bibnamefont {Iwasaki}},\ and\ \bibinfo {author} {\bibfnamefont
  {M.}~\bibnamefont {Hatano}},\ }\href
  {https://doi.org/https://doi.org/10.1016/j.diamond.2019.01.029} {\bibfield
  {journal} {\bibinfo  {journal} {Diam. Relat. Mater.}\ }\textbf {\bibinfo
  {volume} {93}},\ \bibinfo {pages} {75} (\bibinfo {year} {2019})}\BibitemShut
  {NoStop}%
\bibitem [{\citenamefont {Ekimov}\ \emph {et~al.}(2015)\citenamefont {Ekimov},
  \citenamefont {Lyapin}, \citenamefont {Boldyrev}, \citenamefont {Kondrin},
  \citenamefont {Khmelnitskiy}, \citenamefont {Gavva}, \citenamefont
  {Kotereva},\ and\ \citenamefont {Popova}}]{Ekimov2015germanium}%
  \BibitemOpen
  \bibfield  {author} {\bibinfo {author} {\bibfnamefont {E.~A.}\ \bibnamefont
  {Ekimov}}, \bibinfo {author} {\bibfnamefont {S.~G.}\ \bibnamefont {Lyapin}},
  \bibinfo {author} {\bibfnamefont {K.~N.}\ \bibnamefont {Boldyrev}}, \bibinfo
  {author} {\bibfnamefont {M.~V.}\ \bibnamefont {Kondrin}}, \bibinfo {author}
  {\bibfnamefont {R.}~\bibnamefont {Khmelnitskiy}}, \bibinfo {author}
  {\bibfnamefont {V.~A.}\ \bibnamefont {Gavva}}, \bibinfo {author}
  {\bibfnamefont {T.~V.}\ \bibnamefont {Kotereva}},\ and\ \bibinfo {author}
  {\bibfnamefont {M.~N.}\ \bibnamefont {Popova}},\ }\href
  {https://doi.org/10.1134/S0021364015230034} {\bibfield  {journal} {\bibinfo
  {journal} {JETP Lett.}\ }\textbf {\bibinfo {volume} {102}},\ \bibinfo {pages}
  {701} (\bibinfo {year} {2015})}\BibitemShut {NoStop}%
\bibitem [{\citenamefont {Iwasaki}\ \emph {et~al.}(2015)\citenamefont
  {Iwasaki}, \citenamefont {Ishibashi}, \citenamefont {Miyamoto}, \citenamefont
  {Doi}, \citenamefont {Kobayashi}, \citenamefont {Miyazaki}, \citenamefont
  {Tahara}, \citenamefont {Jahnke}, \citenamefont {Rogers}, \citenamefont
  {Naydenov}, \citenamefont {Jelezko}, \citenamefont {Yamasaki}, \citenamefont
  {Nagamachi}, \citenamefont {Inubushi}, \citenamefont {Mizuochi},\ and\
  \citenamefont {Hatano}}]{Iwasaki2015germanium}%
  \BibitemOpen
  \bibfield  {author} {\bibinfo {author} {\bibfnamefont {T.}~\bibnamefont
  {Iwasaki}}, \bibinfo {author} {\bibfnamefont {F.}~\bibnamefont {Ishibashi}},
  \bibinfo {author} {\bibfnamefont {Y.}~\bibnamefont {Miyamoto}}, \bibinfo
  {author} {\bibfnamefont {Y.}~\bibnamefont {Doi}}, \bibinfo {author}
  {\bibfnamefont {S.}~\bibnamefont {Kobayashi}}, \bibinfo {author}
  {\bibfnamefont {T.}~\bibnamefont {Miyazaki}}, \bibinfo {author}
  {\bibfnamefont {K.}~\bibnamefont {Tahara}}, \bibinfo {author} {\bibfnamefont
  {K.~D.}\ \bibnamefont {Jahnke}}, \bibinfo {author} {\bibfnamefont {L.~J.}\
  \bibnamefont {Rogers}}, \bibinfo {author} {\bibfnamefont {B.}~\bibnamefont
  {Naydenov}}, \bibinfo {author} {\bibfnamefont {F.}~\bibnamefont {Jelezko}},
  \bibinfo {author} {\bibfnamefont {S.}~\bibnamefont {Yamasaki}}, \bibinfo
  {author} {\bibfnamefont {S.}~\bibnamefont {Nagamachi}}, \bibinfo {author}
  {\bibfnamefont {T.}~\bibnamefont {Inubushi}}, \bibinfo {author}
  {\bibfnamefont {N.}~\bibnamefont {Mizuochi}},\ and\ \bibinfo {author}
  {\bibfnamefont {M.}~\bibnamefont {Hatano}},\ }\href
  {https://doi.org/10.1038/srep12882} {\bibfield  {journal} {\bibinfo
  {journal} {Sci. Rep.}\ }\textbf {\bibinfo {volume} {5}},\ \bibinfo {pages}
  {12882} (\bibinfo {year} {2015})}\BibitemShut {NoStop}%
\bibitem [{\citenamefont {Ralchenko}\ \emph {et~al.}(2015)\citenamefont
  {Ralchenko}, \citenamefont {Sedov}, \citenamefont {Khomich}, \citenamefont
  {Krivobok}, \citenamefont {Nikolaev}, \citenamefont {Savin}, \citenamefont
  {Vlasov},\ and\ \citenamefont {Konov}}]{Ralchenko2015observation}%
  \BibitemOpen
  \bibfield  {author} {\bibinfo {author} {\bibfnamefont {V.~G.}\ \bibnamefont
  {Ralchenko}}, \bibinfo {author} {\bibfnamefont {V.~S.}\ \bibnamefont
  {Sedov}}, \bibinfo {author} {\bibfnamefont {A.~A.}\ \bibnamefont {Khomich}},
  \bibinfo {author} {\bibfnamefont {V.~S.}\ \bibnamefont {Krivobok}}, \bibinfo
  {author} {\bibfnamefont {S.~N.}\ \bibnamefont {Nikolaev}}, \bibinfo {author}
  {\bibfnamefont {S.}~\bibnamefont {Savin}}, \bibinfo {author} {\bibfnamefont
  {I.~I.}\ \bibnamefont {Vlasov}},\ and\ \bibinfo {author} {\bibfnamefont
  {V.~I.}\ \bibnamefont {Konov}},\ }\href
  {https://doi.org/10.3103/S1068335615060020} {\bibfield  {journal} {\bibinfo
  {journal} {Bull. Leb. Phys. Inst.}\ }\textbf {\bibinfo {volume} {42}},\
  \bibinfo {pages} {165} (\bibinfo {year} {2015})}\BibitemShut {NoStop}%
\bibitem [{\citenamefont {Ekimov}\ \emph {et~al.}(2017)\citenamefont {Ekimov},
  \citenamefont {Krivobok}, \citenamefont {Lyapin}, \citenamefont {Sherin},
  \citenamefont {Gavva},\ and\ \citenamefont
  {Kondrin}}]{Ekimov2017anharmonicity}%
  \BibitemOpen
  \bibfield  {author} {\bibinfo {author} {\bibfnamefont {E.~A.}\ \bibnamefont
  {Ekimov}}, \bibinfo {author} {\bibfnamefont {V.~S.}\ \bibnamefont
  {Krivobok}}, \bibinfo {author} {\bibfnamefont {S.~G.}\ \bibnamefont
  {Lyapin}}, \bibinfo {author} {\bibfnamefont {P.~S.}\ \bibnamefont {Sherin}},
  \bibinfo {author} {\bibfnamefont {V.~A.}\ \bibnamefont {Gavva}},\ and\
  \bibinfo {author} {\bibfnamefont {M.~V.}\ \bibnamefont {Kondrin}},\ }\href
  {https://doi.org/10.1103/PhysRevB.95.094113} {\bibfield  {journal} {\bibinfo
  {journal} {Phys. Rev. B}\ }\textbf {\bibinfo {volume} {95}},\ \bibinfo
  {pages} {094113} (\bibinfo {year} {2017})}\BibitemShut {NoStop}%
\bibitem [{\citenamefont {Palyanov}\ \emph {et~al.}(2016)\citenamefont
  {Palyanov}, \citenamefont {Kupriyanov}, \citenamefont {Borzdov},
  \citenamefont {Khokhryakov},\ and\ \citenamefont
  {Surovtsev}}]{Palyanov2016high}%
  \BibitemOpen
  \bibfield  {author} {\bibinfo {author} {\bibfnamefont {Y.~N.}\ \bibnamefont
  {Palyanov}}, \bibinfo {author} {\bibfnamefont {I.~N.}\ \bibnamefont
  {Kupriyanov}}, \bibinfo {author} {\bibfnamefont {Y.~M.}\ \bibnamefont
  {Borzdov}}, \bibinfo {author} {\bibfnamefont {A.~F.}\ \bibnamefont
  {Khokhryakov}},\ and\ \bibinfo {author} {\bibfnamefont {N.~V.}\ \bibnamefont
  {Surovtsev}},\ }\href {https://doi.org/10.1021/acs.cgd.6b00481} {\bibfield
  {journal} {\bibinfo  {journal} {Cryst. Growth Des.}\ }\textbf {\bibinfo
  {volume} {16}},\ \bibinfo {pages} {3510} (\bibinfo {year}
  {2016})}\BibitemShut {NoStop}%
\bibitem [{\citenamefont {Palyanov}\ \emph {et~al.}(2015)\citenamefont
  {Palyanov}, \citenamefont {Kupriyanov}, \citenamefont {Borzdov},\ and\
  \citenamefont {Surovtsev}}]{Palyanov2015germanium}%
  \BibitemOpen
  \bibfield  {author} {\bibinfo {author} {\bibfnamefont {Y.~N.}\ \bibnamefont
  {Palyanov}}, \bibinfo {author} {\bibfnamefont {I.~N.}\ \bibnamefont
  {Kupriyanov}}, \bibinfo {author} {\bibfnamefont {Y.~M.}\ \bibnamefont
  {Borzdov}},\ and\ \bibinfo {author} {\bibfnamefont {N.~V.}\ \bibnamefont
  {Surovtsev}},\ }\href {https://doi.org/10.1038/srep14789} {\bibfield
  {journal} {\bibinfo  {journal} {Sci. Rep.}\ }\textbf {\bibinfo {volume}
  {5}},\ \bibinfo {pages} {14789} (\bibinfo {year} {2015})}\BibitemShut
  {NoStop}%
\bibitem [{\citenamefont {Bhaskar}\ \emph {et~al.}(2017)\citenamefont
  {Bhaskar}, \citenamefont {Sukachev}, \citenamefont {Sipahigil}, \citenamefont
  {Evans}, \citenamefont {Burek}, \citenamefont {Nguyen}, \citenamefont
  {Rogers}, \citenamefont {Siyushev}, \citenamefont {Metsch}, \citenamefont
  {Park}, \citenamefont {Jelezko}, \citenamefont {Lon\ifmmode~\check{c}\else
  \v{c}\fi{}ar},\ and\ \citenamefont {Lukin}}]{Bhaskar2017quantum}%
  \BibitemOpen
  \bibfield  {author} {\bibinfo {author} {\bibfnamefont {M.~K.}\ \bibnamefont
  {Bhaskar}}, \bibinfo {author} {\bibfnamefont {D.~D.}\ \bibnamefont
  {Sukachev}}, \bibinfo {author} {\bibfnamefont {A.}~\bibnamefont {Sipahigil}},
  \bibinfo {author} {\bibfnamefont {R.~E.}\ \bibnamefont {Evans}}, \bibinfo
  {author} {\bibfnamefont {M.~J.}\ \bibnamefont {Burek}}, \bibinfo {author}
  {\bibfnamefont {C.~T.}\ \bibnamefont {Nguyen}}, \bibinfo {author}
  {\bibfnamefont {L.~J.}\ \bibnamefont {Rogers}}, \bibinfo {author}
  {\bibfnamefont {P.}~\bibnamefont {Siyushev}}, \bibinfo {author}
  {\bibfnamefont {M.~H.}\ \bibnamefont {Metsch}}, \bibinfo {author}
  {\bibfnamefont {H.}~\bibnamefont {Park}}, \bibinfo {author} {\bibfnamefont
  {F.}~\bibnamefont {Jelezko}}, \bibinfo {author} {\bibfnamefont
  {M.}~\bibnamefont {Lon\ifmmode~\check{c}\else \v{c}\fi{}ar}},\ and\ \bibinfo
  {author} {\bibfnamefont {M.~D.}\ \bibnamefont {Lukin}},\ }\href
  {https://doi.org/10.1103/PhysRevLett.118.223603} {\bibfield  {journal}
  {\bibinfo  {journal} {Phys. Rev. Lett.}\ }\textbf {\bibinfo {volume} {118}},\
  \bibinfo {pages} {223603} (\bibinfo {year} {2017})}\BibitemShut {NoStop}%
\bibitem [{\citenamefont {Bray}\ \emph {et~al.}(2018)\citenamefont {Bray},
  \citenamefont {Regan}, \citenamefont {Trycz}, \citenamefont {Previdi},
  \citenamefont {Seniutinas}, \citenamefont {Ganesan}, \citenamefont
  {Kianinia}, \citenamefont {Kim},\ and\ \citenamefont
  {Aharonovich}}]{bray2018single}%
  \BibitemOpen
  \bibfield  {author} {\bibinfo {author} {\bibfnamefont {K.}~\bibnamefont
  {Bray}}, \bibinfo {author} {\bibfnamefont {B.}~\bibnamefont {Regan}},
  \bibinfo {author} {\bibfnamefont {A.}~\bibnamefont {Trycz}}, \bibinfo
  {author} {\bibfnamefont {R.}~\bibnamefont {Previdi}}, \bibinfo {author}
  {\bibfnamefont {G.}~\bibnamefont {Seniutinas}}, \bibinfo {author}
  {\bibfnamefont {K.}~\bibnamefont {Ganesan}}, \bibinfo {author} {\bibfnamefont
  {M.}~\bibnamefont {Kianinia}}, \bibinfo {author} {\bibfnamefont
  {S.}~\bibnamefont {Kim}},\ and\ \bibinfo {author} {\bibfnamefont
  {I.}~\bibnamefont {Aharonovich}},\ }\href
  {https://doi.org/10.1021/acsphotonics.8b00930} {\bibfield  {journal}
  {\bibinfo  {journal} {ACS Photonics}\ }\textbf {\bibinfo {volume} {5}},\
  \bibinfo {pages} {4817} (\bibinfo {year} {2018})}\BibitemShut {NoStop}%
\bibitem [{\citenamefont {Iwasaki}\ \emph {et~al.}(2017)\citenamefont
  {Iwasaki}, \citenamefont {Miyamoto}, \citenamefont {Taniguchi}, \citenamefont
  {Siyushev}, \citenamefont {Metsch}, \citenamefont {Jelezko},\ and\
  \citenamefont {Hatano}}]{Iwasaki2017tin}%
  \BibitemOpen
  \bibfield  {author} {\bibinfo {author} {\bibfnamefont {T.}~\bibnamefont
  {Iwasaki}}, \bibinfo {author} {\bibfnamefont {Y.}~\bibnamefont {Miyamoto}},
  \bibinfo {author} {\bibfnamefont {T.}~\bibnamefont {Taniguchi}}, \bibinfo
  {author} {\bibfnamefont {P.}~\bibnamefont {Siyushev}}, \bibinfo {author}
  {\bibfnamefont {M.~H.}\ \bibnamefont {Metsch}}, \bibinfo {author}
  {\bibfnamefont {F.}~\bibnamefont {Jelezko}},\ and\ \bibinfo {author}
  {\bibfnamefont {M.}~\bibnamefont {Hatano}},\ }\href
  {https://doi.org/10.1103/PhysRevLett.119.253601} {\bibfield  {journal}
  {\bibinfo  {journal} {Phys. Rev. Lett.}\ }\textbf {\bibinfo {volume} {119}},\
  \bibinfo {pages} {253601} (\bibinfo {year} {2017})}\BibitemShut {NoStop}%
\bibitem [{\citenamefont {Ekimov}\ \emph {et~al.}(2018)\citenamefont {Ekimov},
  \citenamefont {Lyapin},\ and\ \citenamefont {Kondrin}}]{Ekimov2018tin}%
  \BibitemOpen
  \bibfield  {author} {\bibinfo {author} {\bibfnamefont {E.}~\bibnamefont
  {Ekimov}}, \bibinfo {author} {\bibfnamefont {S.}~\bibnamefont {Lyapin}},\
  and\ \bibinfo {author} {\bibfnamefont {M.}~\bibnamefont {Kondrin}},\ }\href
  {https://doi.org/https://doi.org/10.1016/j.diamond.2018.06.014} {\bibfield
  {journal} {\bibinfo  {journal} {Diam. Relat. Mater.}\ }\textbf {\bibinfo
  {volume} {87}},\ \bibinfo {pages} {223} (\bibinfo {year} {2018})}\BibitemShut
  {NoStop}%
\bibitem [{\citenamefont {Ditalia~Tchernij}\ \emph {et~al.}(2017)\citenamefont
  {Ditalia~Tchernij}, \citenamefont {Herzig}, \citenamefont {Forneris},
  \citenamefont {K\"upper}, \citenamefont {Pezzagna}, \citenamefont {Traina},
  \citenamefont {Moreva}, \citenamefont {Degiovanni}, \citenamefont {Brida},
  \citenamefont {Skukan}, \citenamefont {Genovese}, \citenamefont
  {Jak\ifmmode~\breve{s}\else \u{s}\fi{}i\'c}, \citenamefont {Meijer},\ and\
  \citenamefont {Olivero}}]{tchernij2017single}%
  \BibitemOpen
  \bibfield  {author} {\bibinfo {author} {\bibfnamefont {S.}~\bibnamefont
  {Ditalia~Tchernij}}, \bibinfo {author} {\bibfnamefont {T.}~\bibnamefont
  {Herzig}}, \bibinfo {author} {\bibfnamefont {J.}~\bibnamefont {Forneris}},
  \bibinfo {author} {\bibfnamefont {J.}~\bibnamefont {K\"upper}}, \bibinfo
  {author} {\bibfnamefont {S.}~\bibnamefont {Pezzagna}}, \bibinfo {author}
  {\bibfnamefont {P.}~\bibnamefont {Traina}}, \bibinfo {author} {\bibfnamefont
  {E.}~\bibnamefont {Moreva}}, \bibinfo {author} {\bibfnamefont {I.~P.}\
  \bibnamefont {Degiovanni}}, \bibinfo {author} {\bibfnamefont
  {G.}~\bibnamefont {Brida}}, \bibinfo {author} {\bibfnamefont
  {N.}~\bibnamefont {Skukan}}, \bibinfo {author} {\bibfnamefont
  {M.}~\bibnamefont {Genovese}}, \bibinfo {author} {\bibfnamefont
  {M.}~\bibnamefont {Jak\ifmmode~\breve{s}\else \u{s}\fi{}i\'c}}, \bibinfo
  {author} {\bibfnamefont {J.}~\bibnamefont {Meijer}},\ and\ \bibinfo {author}
  {\bibfnamefont {P.}~\bibnamefont {Olivero}},\ }\href
  {https://doi.org/10.1021/acsphotonics.7b00904} {\bibfield  {journal}
  {\bibinfo  {journal} {ACS Photonics}\ }\textbf {\bibinfo {volume} {4}},\
  \bibinfo {pages} {2580} (\bibinfo {year} {2017})}\BibitemShut {NoStop}%
\bibitem [{\citenamefont {Palyanov}\ \emph {et~al.}(2019)\citenamefont
  {Palyanov}, \citenamefont {Kupriyanov},\ and\ \citenamefont
  {Borzdov}}]{palyanov2019high}%
  \BibitemOpen
  \bibfield  {author} {\bibinfo {author} {\bibfnamefont {Y.~N.}\ \bibnamefont
  {Palyanov}}, \bibinfo {author} {\bibfnamefont {I.~N.}\ \bibnamefont
  {Kupriyanov}},\ and\ \bibinfo {author} {\bibfnamefont {Y.~M.}\ \bibnamefont
  {Borzdov}},\ }\href
  {https://doi.org/https://doi.org/10.1016/j.carbon.2018.11.084} {\bibfield
  {journal} {\bibinfo  {journal} {Carbon}\ }\textbf {\bibinfo {volume} {143}},\
  \bibinfo {pages} {769} (\bibinfo {year} {2019})}\BibitemShut {NoStop}%
\bibitem [{\citenamefont {Alkahtani}\ \emph {et~al.}(2018)\citenamefont
  {Alkahtani}, \citenamefont {Cojocaru}, \citenamefont {Liu}, \citenamefont
  {Herzig}, \citenamefont {Meijer}, \citenamefont {K{\"u}pper}, \citenamefont
  {L{\"u}hmann}, \citenamefont {Akimov},\ and\ \citenamefont
  {Hemmer}}]{Alkahtani2018tin}%
  \BibitemOpen
  \bibfield  {author} {\bibinfo {author} {\bibfnamefont {M.}~\bibnamefont
  {Alkahtani}}, \bibinfo {author} {\bibfnamefont {I.}~\bibnamefont {Cojocaru}},
  \bibinfo {author} {\bibfnamefont {X.}~\bibnamefont {Liu}}, \bibinfo {author}
  {\bibfnamefont {T.}~\bibnamefont {Herzig}}, \bibinfo {author} {\bibfnamefont
  {J.}~\bibnamefont {Meijer}}, \bibinfo {author} {\bibfnamefont
  {J.}~\bibnamefont {K{\"u}pper}}, \bibinfo {author} {\bibfnamefont
  {T.}~\bibnamefont {L{\"u}hmann}}, \bibinfo {author} {\bibfnamefont {A.~V.}\
  \bibnamefont {Akimov}},\ and\ \bibinfo {author} {\bibfnamefont {P.~R.}\
  \bibnamefont {Hemmer}},\ }\href {https://doi.org/10.1063/1.5037053}
  {\bibfield  {journal} {\bibinfo  {journal} {Appl. Phys. Lett.}\ }\textbf
  {\bibinfo {volume} {112}},\ \bibinfo {pages} {241902} (\bibinfo {year}
  {2018})}\BibitemShut {NoStop}%
\bibitem [{\citenamefont {Rugar}\ \emph {et~al.}(2019)\citenamefont {Rugar},
  \citenamefont {Dory}, \citenamefont {Sun},\ and\ \citenamefont {Vu\ifmmode
  \check{c}\else \v{c}\fi{}kovi\ifmmode~\acute{c}\else
  \'{c}\fi{}}}]{rugar2019char}%
  \BibitemOpen
  \bibfield  {author} {\bibinfo {author} {\bibfnamefont {A.~E.}\ \bibnamefont
  {Rugar}}, \bibinfo {author} {\bibfnamefont {C.}~\bibnamefont {Dory}},
  \bibinfo {author} {\bibfnamefont {S.}~\bibnamefont {Sun}},\ and\ \bibinfo
  {author} {\bibfnamefont {J.}~\bibnamefont {Vu\ifmmode \check{c}\else
  \v{c}\fi{}kovi\ifmmode~\acute{c}\else \'{c}\fi{}}},\ }\href
  {https://doi.org/10.1103/PhysRevB.99.205417} {\bibfield  {journal} {\bibinfo
  {journal} {Phys. Rev. B}\ }\textbf {\bibinfo {volume} {99}},\ \bibinfo
  {pages} {205417} (\bibinfo {year} {2019})}\BibitemShut {NoStop}%
\bibitem [{\citenamefont {Wahl}\ \emph {et~al.}(2020)\citenamefont {Wahl},
  \citenamefont {Correia}, \citenamefont {Villarreal}, \citenamefont
  {Bourgeois}, \citenamefont {Gulka}, \citenamefont {Nesl{\'a}dek},
  \citenamefont {Vantomme},\ and\ \citenamefont {Pereira}}]{wahl2020direct}%
  \BibitemOpen
  \bibfield  {author} {\bibinfo {author} {\bibfnamefont {U.}~\bibnamefont
  {Wahl}}, \bibinfo {author} {\bibfnamefont {J.~G.}\ \bibnamefont {Correia}},
  \bibinfo {author} {\bibfnamefont {R.}~\bibnamefont {Villarreal}}, \bibinfo
  {author} {\bibfnamefont {E.}~\bibnamefont {Bourgeois}}, \bibinfo {author}
  {\bibfnamefont {M.}~\bibnamefont {Gulka}}, \bibinfo {author} {\bibfnamefont
  {M.}~\bibnamefont {Nesl{\'a}dek}}, \bibinfo {author} {\bibfnamefont
  {A.}~\bibnamefont {Vantomme}},\ and\ \bibinfo {author} {\bibfnamefont
  {L.~M.~C.}\ \bibnamefont {Pereira}},\ }\href
  {https://link.aps.org/doi/10.1103/PhysRevLett.125.045301} {\bibfield
  {journal} {\bibinfo  {journal} {Phys. Rev. Lett.}\ }\textbf {\bibinfo
  {volume} {125}},\ \bibinfo {pages} {045301} (\bibinfo {year}
  {2020})}\BibitemShut {NoStop}%
\bibitem [{\citenamefont {Fukuta}\ \emph {et~al.}(2021)\citenamefont {Fukuta},
  \citenamefont {Murakami}, \citenamefont {Ohfuji}, \citenamefont {Shinmei},
  \citenamefont {Irifune},\ and\ \citenamefont {Ishikawa}}]{fukuta2021sn}%
  \BibitemOpen
  \bibfield  {author} {\bibinfo {author} {\bibfnamefont {R.}~\bibnamefont
  {Fukuta}}, \bibinfo {author} {\bibfnamefont {Y.}~\bibnamefont {Murakami}},
  \bibinfo {author} {\bibfnamefont {H.}~\bibnamefont {Ohfuji}}, \bibinfo
  {author} {\bibfnamefont {T.}~\bibnamefont {Shinmei}}, \bibinfo {author}
  {\bibfnamefont {T.}~\bibnamefont {Irifune}},\ and\ \bibinfo {author}
  {\bibfnamefont {F.}~\bibnamefont {Ishikawa}},\ }\href
  {https://doi.org/10.35848/1347-4065/abdc31} {\bibfield  {journal} {\bibinfo
  {journal} {Japanese Journal of Applied Physics}\ }\textbf {\bibinfo {volume}
  {60}},\ \bibinfo {pages} {035501} (\bibinfo {year} {2021})}\BibitemShut
  {NoStop}%
\bibitem [{\citenamefont {G\"{o}rlitz}\ \emph {et~al.}(2020)\citenamefont
  {G\"{o}rlitz}, \citenamefont {Herrmann}, \citenamefont {Thiering},
  \citenamefont {Fuchs}, \citenamefont {Gandil}, \citenamefont {Iwasaki},
  \citenamefont {Taniguchi}, \citenamefont {Kieschnick}, \citenamefont
  {Meijer}, \citenamefont {Hatano}, \citenamefont {Gali},\ and\ \citenamefont
  {Becher}}]{Gorlitz2020spectroscopic}%
  \BibitemOpen
  \bibfield  {author} {\bibinfo {author} {\bibfnamefont {J.}~\bibnamefont
  {G\"{o}rlitz}}, \bibinfo {author} {\bibfnamefont {D.}~\bibnamefont
  {Herrmann}}, \bibinfo {author} {\bibfnamefont {G.}~\bibnamefont {Thiering}},
  \bibinfo {author} {\bibfnamefont {P.}~\bibnamefont {Fuchs}}, \bibinfo
  {author} {\bibfnamefont {M.}~\bibnamefont {Gandil}}, \bibinfo {author}
  {\bibfnamefont {T.}~\bibnamefont {Iwasaki}}, \bibinfo {author} {\bibfnamefont
  {T.}~\bibnamefont {Taniguchi}}, \bibinfo {author} {\bibfnamefont
  {M.}~\bibnamefont {Kieschnick}}, \bibinfo {author} {\bibfnamefont
  {J.}~\bibnamefont {Meijer}}, \bibinfo {author} {\bibfnamefont
  {M.}~\bibnamefont {Hatano}}, \bibinfo {author} {\bibfnamefont
  {A.}~\bibnamefont {Gali}},\ and\ \bibinfo {author} {\bibfnamefont
  {C.}~\bibnamefont {Becher}},\ }\href
  {https://doi.org/10.1088/1367-2630/ab6631} {\bibfield  {journal} {\bibinfo
  {journal} {New Journal of Physics}\ }\textbf {\bibinfo {volume} {22}},\
  \bibinfo {pages} {013048} (\bibinfo {year} {2020})}\BibitemShut {NoStop}%
\bibitem [{\citenamefont {Trusheim}\ \emph {et~al.}(2020)\citenamefont
  {Trusheim}, \citenamefont {Pingault}, \citenamefont {Wan}, \citenamefont
  {G\"undo\ifmmode~\breve{g}\else \u{g}\fi{}an}, \citenamefont {De~Santis},
  \citenamefont {Debroux}, \citenamefont {Gangloff}, \citenamefont {Purser},
  \citenamefont {Chen}, \citenamefont {Walsh}, \citenamefont {Rose},
  \citenamefont {Becker}, \citenamefont {Lienhard}, \citenamefont {Bersin},
  \citenamefont {Paradeisanos}, \citenamefont {Wang}, \citenamefont {Lyzwa},
  \citenamefont {Montblanch}, \citenamefont {Malladi}, \citenamefont {Bakhru},
  \citenamefont {Ferrari}, \citenamefont {Walmsley}, \citenamefont
  {Atat\"ure},\ and\ \citenamefont {Englund}}]{Trusheim2020transform}%
  \BibitemOpen
  \bibfield  {author} {\bibinfo {author} {\bibfnamefont {M.~E.}\ \bibnamefont
  {Trusheim}}, \bibinfo {author} {\bibfnamefont {B.}~\bibnamefont {Pingault}},
  \bibinfo {author} {\bibfnamefont {N.~H.}\ \bibnamefont {Wan}}, \bibinfo
  {author} {\bibfnamefont {M.}~\bibnamefont {G\"undo\ifmmode~\breve{g}\else
  \u{g}\fi{}an}}, \bibinfo {author} {\bibfnamefont {L.}~\bibnamefont
  {De~Santis}}, \bibinfo {author} {\bibfnamefont {R.}~\bibnamefont {Debroux}},
  \bibinfo {author} {\bibfnamefont {D.}~\bibnamefont {Gangloff}}, \bibinfo
  {author} {\bibfnamefont {C.}~\bibnamefont {Purser}}, \bibinfo {author}
  {\bibfnamefont {K.~C.}\ \bibnamefont {Chen}}, \bibinfo {author}
  {\bibfnamefont {M.}~\bibnamefont {Walsh}}, \bibinfo {author} {\bibfnamefont
  {J.~J.}\ \bibnamefont {Rose}}, \bibinfo {author} {\bibfnamefont {J.~N.}\
  \bibnamefont {Becker}}, \bibinfo {author} {\bibfnamefont {B.}~\bibnamefont
  {Lienhard}}, \bibinfo {author} {\bibfnamefont {E.}~\bibnamefont {Bersin}},
  \bibinfo {author} {\bibfnamefont {I.}~\bibnamefont {Paradeisanos}}, \bibinfo
  {author} {\bibfnamefont {G.}~\bibnamefont {Wang}}, \bibinfo {author}
  {\bibfnamefont {D.}~\bibnamefont {Lyzwa}}, \bibinfo {author} {\bibfnamefont
  {A.~R.-P.}\ \bibnamefont {Montblanch}}, \bibinfo {author} {\bibfnamefont
  {G.}~\bibnamefont {Malladi}}, \bibinfo {author} {\bibfnamefont
  {H.}~\bibnamefont {Bakhru}}, \bibinfo {author} {\bibfnamefont {A.~C.}\
  \bibnamefont {Ferrari}}, \bibinfo {author} {\bibfnamefont {I.~A.}\
  \bibnamefont {Walmsley}}, \bibinfo {author} {\bibfnamefont {M.}~\bibnamefont
  {Atat\"ure}},\ and\ \bibinfo {author} {\bibfnamefont {D.}~\bibnamefont
  {Englund}},\ }\href {https://doi.org/10.1103/PhysRevLett.124.023602}
  {\bibfield  {journal} {\bibinfo  {journal} {Phys. Rev. Lett.}\ }\textbf
  {\bibinfo {volume} {124}},\ \bibinfo {pages} {023602} (\bibinfo {year}
  {2020})}\BibitemShut {NoStop}%
\bibitem [{\citenamefont {Rugar}\ \emph {et~al.}(2021)\citenamefont {Rugar},
  \citenamefont {Aghaeimeibodi}, \citenamefont {Riedel}, \citenamefont {Dory},
  \citenamefont {Lu}, \citenamefont {McQuade}, \citenamefont {Shen},
  \citenamefont {Melosh},\ and\ \citenamefont {Vu\ifmmode \check{c}\else
  \v{c}\fi{}kovi\ifmmode~\acute{c}\else \'{c}\fi{}}}]{Rugar2021quantum}%
  \BibitemOpen
  \bibfield  {author} {\bibinfo {author} {\bibfnamefont {A.~E.}\ \bibnamefont
  {Rugar}}, \bibinfo {author} {\bibfnamefont {S.}~\bibnamefont
  {Aghaeimeibodi}}, \bibinfo {author} {\bibfnamefont {D.}~\bibnamefont
  {Riedel}}, \bibinfo {author} {\bibfnamefont {C.}~\bibnamefont {Dory}},
  \bibinfo {author} {\bibfnamefont {H.}~\bibnamefont {Lu}}, \bibinfo {author}
  {\bibfnamefont {P.~J.}\ \bibnamefont {McQuade}}, \bibinfo {author}
  {\bibfnamefont {Z.-X.}\ \bibnamefont {Shen}}, \bibinfo {author}
  {\bibfnamefont {N.~A.}\ \bibnamefont {Melosh}},\ and\ \bibinfo {author}
  {\bibfnamefont {J.}~\bibnamefont {Vu\ifmmode \check{c}\else
  \v{c}\fi{}kovi\ifmmode~\acute{c}\else \'{c}\fi{}}},\ }\href
  {https://doi.org/10.1103/PhysRevX.11.031021} {\bibfield  {journal} {\bibinfo
  {journal} {Phys. Rev. X}\ }\textbf {\bibinfo {volume} {11}},\ \bibinfo
  {pages} {031021} (\bibinfo {year} {2021})}\BibitemShut {NoStop}%
\bibitem [{\citenamefont {Trusheim}\ \emph {et~al.}(2019)\citenamefont
  {Trusheim}, \citenamefont {Wan}, \citenamefont {Chen}, \citenamefont
  {Ciccarino}, \citenamefont {Flick}, \citenamefont {Sundararaman},
  \citenamefont {Malladi}, \citenamefont {Bersin}, \citenamefont {Walsh},
  \citenamefont {Lienhard}, \citenamefont {Bakhru}, \citenamefont {Narang},\
  and\ \citenamefont {Englund}}]{Trusheim2019lead}%
  \BibitemOpen
  \bibfield  {author} {\bibinfo {author} {\bibfnamefont {M.~E.}\ \bibnamefont
  {Trusheim}}, \bibinfo {author} {\bibfnamefont {N.~H.}\ \bibnamefont {Wan}},
  \bibinfo {author} {\bibfnamefont {K.~C.}\ \bibnamefont {Chen}}, \bibinfo
  {author} {\bibfnamefont {C.~J.}\ \bibnamefont {Ciccarino}}, \bibinfo {author}
  {\bibfnamefont {J.}~\bibnamefont {Flick}}, \bibinfo {author} {\bibfnamefont
  {R.}~\bibnamefont {Sundararaman}}, \bibinfo {author} {\bibfnamefont
  {G.}~\bibnamefont {Malladi}}, \bibinfo {author} {\bibfnamefont
  {E.}~\bibnamefont {Bersin}}, \bibinfo {author} {\bibfnamefont
  {M.}~\bibnamefont {Walsh}}, \bibinfo {author} {\bibfnamefont
  {B.}~\bibnamefont {Lienhard}}, \bibinfo {author} {\bibfnamefont
  {H.}~\bibnamefont {Bakhru}}, \bibinfo {author} {\bibfnamefont
  {P.}~\bibnamefont {Narang}},\ and\ \bibinfo {author} {\bibfnamefont
  {D.}~\bibnamefont {Englund}},\ }\href
  {https://doi.org/10.1103/PhysRevB.99.075430} {\bibfield  {journal} {\bibinfo
  {journal} {Phys. Rev. B}\ }\textbf {\bibinfo {volume} {99}},\ \bibinfo
  {pages} {075430} (\bibinfo {year} {2019})}\BibitemShut {NoStop}%
\bibitem [{\citenamefont {Ditalia~Tchernij}\ \emph {et~al.}(2018)\citenamefont
  {Ditalia~Tchernij}, \citenamefont {L\"{u}hmann}, \citenamefont {Herzig},
  \citenamefont {K\"{u}pper}, \citenamefont {Damin}, \citenamefont
  {Santonocito}, \citenamefont {Signorile}, \citenamefont {Traina},
  \citenamefont {Moreva}, \citenamefont {Celegato}, \citenamefont {Pezzagna},
  \citenamefont {Degiovanni}, \citenamefont {Olivero}, \citenamefont
  {Jak\ifmmode~\breve{s}\else \u{s}\fi{}i\'c}, \citenamefont {Meijer},
  \citenamefont {Genovese},\ and\ \citenamefont
  {Forneris}}]{tchernij2018single}%
  \BibitemOpen
  \bibfield  {author} {\bibinfo {author} {\bibfnamefont {S.}~\bibnamefont
  {Ditalia~Tchernij}}, \bibinfo {author} {\bibfnamefont {T.}~\bibnamefont
  {L\"{u}hmann}}, \bibinfo {author} {\bibfnamefont {T.}~\bibnamefont {Herzig}},
  \bibinfo {author} {\bibfnamefont {J.}~\bibnamefont {K\"{u}pper}}, \bibinfo
  {author} {\bibfnamefont {A.}~\bibnamefont {Damin}}, \bibinfo {author}
  {\bibfnamefont {S.}~\bibnamefont {Santonocito}}, \bibinfo {author}
  {\bibfnamefont {M.}~\bibnamefont {Signorile}}, \bibinfo {author}
  {\bibfnamefont {P.}~\bibnamefont {Traina}}, \bibinfo {author} {\bibfnamefont
  {E.}~\bibnamefont {Moreva}}, \bibinfo {author} {\bibfnamefont
  {F.}~\bibnamefont {Celegato}}, \bibinfo {author} {\bibfnamefont
  {S.}~\bibnamefont {Pezzagna}}, \bibinfo {author} {\bibfnamefont {I.~P.}\
  \bibnamefont {Degiovanni}}, \bibinfo {author} {\bibfnamefont
  {P.}~\bibnamefont {Olivero}}, \bibinfo {author} {\bibfnamefont
  {M.}~\bibnamefont {Jak\ifmmode~\breve{s}\else \u{s}\fi{}i\'c}}, \bibinfo
  {author} {\bibfnamefont {J.}~\bibnamefont {Meijer}}, \bibinfo {author}
  {\bibfnamefont {P.~M.}\ \bibnamefont {Genovese}},\ and\ \bibinfo {author}
  {\bibfnamefont {J.}~\bibnamefont {Forneris}},\ }\href
  {https://doi.org/10.1021/acsphotonics.8b01013} {\bibfield  {journal}
  {\bibinfo  {journal} {ACS Photonics}\ }\textbf {\bibinfo {volume} {5}},\
  \bibinfo {pages} {4864} (\bibinfo {year} {2018})}\BibitemShut {NoStop}%
\bibitem [{\citenamefont {Castelletto}\ \emph {et~al.}(2014)\citenamefont
  {Castelletto}, \citenamefont {Johnson}, \citenamefont {Iv{\'a}dy},
  \citenamefont {Stavrias}, \citenamefont {Umeda}, \citenamefont {Gali},\ and\
  \citenamefont {Ohshima}}]{Castelletto2014a}%
  \BibitemOpen
  \bibfield  {author} {\bibinfo {author} {\bibfnamefont {S.}~\bibnamefont
  {Castelletto}}, \bibinfo {author} {\bibfnamefont {B.~C.}\ \bibnamefont
  {Johnson}}, \bibinfo {author} {\bibfnamefont {V.}~\bibnamefont {Iv{\'a}dy}},
  \bibinfo {author} {\bibfnamefont {N.}~\bibnamefont {Stavrias}}, \bibinfo
  {author} {\bibfnamefont {T.}~\bibnamefont {Umeda}}, \bibinfo {author}
  {\bibfnamefont {A.}~\bibnamefont {Gali}},\ and\ \bibinfo {author}
  {\bibfnamefont {T.}~\bibnamefont {Ohshima}},\ }\href
  {http://dx.doi.org/10.1038/nmat3806} {\bibfield  {journal} {\bibinfo
  {journal} {Nat. Mater.}\ }\textbf {\bibinfo {volume} {13}},\ \bibinfo {pages}
  {151} (\bibinfo {year} {2014})}\BibitemShut {NoStop}%
\bibitem [{\citenamefont {Bockstedte}\ \emph {et~al.}(2004)\citenamefont
  {Bockstedte}, \citenamefont {Mattausch},\ and\ \citenamefont
  {Pankratov}}]{Bockstedte2004ab}%
  \BibitemOpen
  \bibfield  {author} {\bibinfo {author} {\bibfnamefont {M.}~\bibnamefont
  {Bockstedte}}, \bibinfo {author} {\bibfnamefont {A.}~\bibnamefont
  {Mattausch}},\ and\ \bibinfo {author} {\bibfnamefont {O.}~\bibnamefont
  {Pankratov}},\ }\href {https://doi.org/10.1103/PhysRevB.69.235202} {\bibfield
   {journal} {\bibinfo  {journal} {Phys. Rev. B}\ }\textbf {\bibinfo {volume}
  {69}},\ \bibinfo {pages} {235202} (\bibinfo {year} {2004})}\BibitemShut
  {NoStop}%
\bibitem [{\citenamefont {Kimoto}\ and\ \citenamefont
  {Cooper}(2014)}]{Kimoto2014fundamentals}%
  \BibitemOpen
  \bibfield  {author} {\bibinfo {author} {\bibfnamefont {T.}~\bibnamefont
  {Kimoto}}\ and\ \bibinfo {author} {\bibfnamefont {J.}~\bibnamefont
  {Cooper}},\ }\href {https://books.google.com/books?id=WJzmBQAAQBAJ} {\emph
  {\bibinfo {title} {Fundamentals of Silicon Carbide Technology: Growth,
  Characterization, Devices and Applications}}},\ Wiley - IEEE\ (\bibinfo
  {publisher} {Wiley},\ \bibinfo {year} {2014})\BibitemShut {NoStop}%
\bibitem [{\citenamefont {Kuate~Defo}\ \emph {et~al.}(2018)\citenamefont
  {Kuate~Defo}, \citenamefont {Zhang}, \citenamefont {Bracher}, \citenamefont
  {Kim}, \citenamefont {Hu},\ and\ \citenamefont
  {Kaxiras}}]{Kuate2018energetics}%
  \BibitemOpen
  \bibfield  {author} {\bibinfo {author} {\bibfnamefont {R.}~\bibnamefont
  {Kuate~Defo}}, \bibinfo {author} {\bibfnamefont {X.}~\bibnamefont {Zhang}},
  \bibinfo {author} {\bibfnamefont {D.}~\bibnamefont {Bracher}}, \bibinfo
  {author} {\bibfnamefont {G.}~\bibnamefont {Kim}}, \bibinfo {author}
  {\bibfnamefont {E.}~\bibnamefont {Hu}},\ and\ \bibinfo {author}
  {\bibfnamefont {E.}~\bibnamefont {Kaxiras}},\ }\href
  {https://doi.org/10.1103/PhysRevB.98.104103} {\bibfield  {journal} {\bibinfo
  {journal} {Phys. Rev. B}\ }\textbf {\bibinfo {volume} {98}},\ \bibinfo
  {pages} {104103} (\bibinfo {year} {2018})}\BibitemShut {NoStop}%
\bibitem [{\citenamefont {Gadalla}\ \emph {et~al.}(2021)\citenamefont
  {Gadalla}, \citenamefont {Greenspon}, \citenamefont {Kuate~Defo},
  \citenamefont {Zhang},\ and\ \citenamefont {Hu}}]{Gadalla2021enhanced}%
  \BibitemOpen
  \bibfield  {author} {\bibinfo {author} {\bibfnamefont {M.~N.}\ \bibnamefont
  {Gadalla}}, \bibinfo {author} {\bibfnamefont {A.~S.}\ \bibnamefont
  {Greenspon}}, \bibinfo {author} {\bibfnamefont {R.}~\bibnamefont
  {Kuate~Defo}}, \bibinfo {author} {\bibfnamefont {X.}~\bibnamefont {Zhang}},\
  and\ \bibinfo {author} {\bibfnamefont {E.~L.}\ \bibnamefont {Hu}},\ }\href
  {https://doi.org/10.1073/pnas.2021768118} {\bibfield  {journal} {\bibinfo
  {journal} {Proceedings of the National Academy of Sciences}\ }\textbf
  {\bibinfo {volume} {118}},\ \bibinfo {pages} {e2021768118} (\bibinfo {year}
  {2021})}\BibitemShut {NoStop}%
\bibitem [{\citenamefont {{Kuate Defo}}\ \emph {et~al.}(2019)\citenamefont
  {{Kuate Defo}}, \citenamefont {Wang},\ and\ \citenamefont
  {Manjunathaiah}}]{Kuate2019parallel}%
  \BibitemOpen
  \bibfield  {author} {\bibinfo {author} {\bibfnamefont {R.}~\bibnamefont
  {{Kuate Defo}}}, \bibinfo {author} {\bibfnamefont {R.}~\bibnamefont {Wang}},\
  and\ \bibinfo {author} {\bibfnamefont {M.}~\bibnamefont {Manjunathaiah}},\
  }\href {https://doi.org/https://doi.org/10.1016/j.jocs.2019.07.005}
  {\bibfield  {journal} {\bibinfo  {journal} {Journal of Computational
  Science}\ }\textbf {\bibinfo {volume} {36}},\ \bibinfo {pages} {101018}
  (\bibinfo {year} {2019})}\BibitemShut {NoStop}%
\bibitem [{\citenamefont {Bracher}\ \emph {et~al.}(2017)\citenamefont
  {Bracher}, \citenamefont {Zhang},\ and\ \citenamefont
  {Hu}}]{Bracher2017selective}%
  \BibitemOpen
  \bibfield  {author} {\bibinfo {author} {\bibfnamefont {D.~O.}\ \bibnamefont
  {Bracher}}, \bibinfo {author} {\bibfnamefont {X.}~\bibnamefont {Zhang}},\
  and\ \bibinfo {author} {\bibfnamefont {E.~L.}\ \bibnamefont {Hu}},\ }\href
  {https://doi.org/10.1073/pnas.1704219114} {\bibfield  {journal} {\bibinfo
  {journal} {Proceedings of the National Academy of Sciences}\ }\textbf
  {\bibinfo {volume} {114}},\ \bibinfo {pages} {4060} (\bibinfo {year}
  {2017})}\BibitemShut {NoStop}%
\bibitem [{\citenamefont {Soykal}\ \emph {et~al.}(2016)\citenamefont {Soykal},
  \citenamefont {Dev},\ and\ \citenamefont {Economou}}]{Soykal2016silicon}%
  \BibitemOpen
  \bibfield  {author} {\bibinfo {author} {\bibfnamefont {O.~O.}\ \bibnamefont
  {Soykal}}, \bibinfo {author} {\bibfnamefont {P.}~\bibnamefont {Dev}},\ and\
  \bibinfo {author} {\bibfnamefont {S.~E.}\ \bibnamefont {Economou}},\ }\href
  {https://doi.org/10.1103/PhysRevB.93.081207} {\bibfield  {journal} {\bibinfo
  {journal} {Phys. Rev. B}\ }\textbf {\bibinfo {volume} {93}},\ \bibinfo
  {pages} {081207(R)} (\bibinfo {year} {2016})}\BibitemShut {NoStop}%
\bibitem [{\citenamefont {Soykal}\ and\ \citenamefont
  {Reinecke}(2017)}]{Soykal2017quantum}%
  \BibitemOpen
  \bibfield  {author} {\bibinfo {author} {\bibfnamefont {O.~O.}\ \bibnamefont
  {Soykal}}\ and\ \bibinfo {author} {\bibfnamefont {T.~L.}\ \bibnamefont
  {Reinecke}},\ }\href {https://doi.org/10.1103/PhysRevB.95.081405} {\bibfield
  {journal} {\bibinfo  {journal} {Phys. Rev. B}\ }\textbf {\bibinfo {volume}
  {95}},\ \bibinfo {pages} {081405(R)} (\bibinfo {year} {2017})}\BibitemShut
  {NoStop}%
\bibitem [{\citenamefont {Gali}(2011)}]{Gali2011time}%
  \BibitemOpen
  \bibfield  {author} {\bibinfo {author} {\bibfnamefont {A.}~\bibnamefont
  {Gali}},\ }\href {https://doi.org/10.1002/pssb.201046254} {\bibfield
  {journal} {\bibinfo  {journal} {physica status solidi (b)}\ }\textbf
  {\bibinfo {volume} {248}},\ \bibinfo {pages} {1337} (\bibinfo {year}
  {2011})}\BibitemShut {NoStop}%
\bibitem [{\citenamefont {Awschalom}\ \emph {et~al.}(2018)\citenamefont
  {Awschalom}, \citenamefont {Hanson}, \citenamefont {Wrachtrup},\ and\
  \citenamefont {Zhou}}]{Awschalom2018quantum}%
  \BibitemOpen
  \bibfield  {author} {\bibinfo {author} {\bibfnamefont {D.~D.}\ \bibnamefont
  {Awschalom}}, \bibinfo {author} {\bibfnamefont {R.}~\bibnamefont {Hanson}},
  \bibinfo {author} {\bibfnamefont {J.}~\bibnamefont {Wrachtrup}},\ and\
  \bibinfo {author} {\bibfnamefont {B.~B.}\ \bibnamefont {Zhou}},\ }\href
  {https://doi.org/10.1038/s41566-018-0232-2} {\bibfield  {journal} {\bibinfo
  {journal} {Nature Photonics}\ }\textbf {\bibinfo {volume} {12}},\ \bibinfo
  {pages} {516} (\bibinfo {year} {2018})}\BibitemShut {NoStop}%
\bibitem [{\citenamefont {Wolfowicz}\ \emph {et~al.}(2021)\citenamefont
  {Wolfowicz}, \citenamefont {Heremans}, \citenamefont {Anderson},
  \citenamefont {Kanai}, \citenamefont {Seo}, \citenamefont {Gali},
  \citenamefont {Galli},\ and\ \citenamefont
  {Awschalom}}]{Wolfowicz2021quantum}%
  \BibitemOpen
  \bibfield  {author} {\bibinfo {author} {\bibfnamefont {G.}~\bibnamefont
  {Wolfowicz}}, \bibinfo {author} {\bibfnamefont {F.~J.}\ \bibnamefont
  {Heremans}}, \bibinfo {author} {\bibfnamefont {C.~P.}\ \bibnamefont
  {Anderson}}, \bibinfo {author} {\bibfnamefont {S.}~\bibnamefont {Kanai}},
  \bibinfo {author} {\bibfnamefont {H.}~\bibnamefont {Seo}}, \bibinfo {author}
  {\bibfnamefont {A.}~\bibnamefont {Gali}}, \bibinfo {author} {\bibfnamefont
  {G.}~\bibnamefont {Galli}},\ and\ \bibinfo {author} {\bibfnamefont {D.~D.}\
  \bibnamefont {Awschalom}},\ }\href
  {https://doi.org/10.1038/s41578-021-00306-y} {\bibfield  {journal} {\bibinfo
  {journal} {Nature Reviews Materials}\ }\textbf {\bibinfo {volume} {6}},\
  \bibinfo {pages} {906} (\bibinfo {year} {2021})}\BibitemShut {NoStop}%
\bibitem [{\citenamefont {Whiteley}\ \emph {et~al.}(2019)\citenamefont
  {Whiteley}, \citenamefont {Wolfowicz}, \citenamefont {Anderson},
  \citenamefont {Bourassa}, \citenamefont {Ma}, \citenamefont {Ye},
  \citenamefont {Koolstra}, \citenamefont {Satzinger}, \citenamefont {Holt},
  \citenamefont {Heremans}, \citenamefont {Cleland}, \citenamefont {Schuster},
  \citenamefont {Galli},\ and\ \citenamefont {Awschalom}}]{Whiteley2019spin}%
  \BibitemOpen
  \bibfield  {author} {\bibinfo {author} {\bibfnamefont {S.~J.}\ \bibnamefont
  {Whiteley}}, \bibinfo {author} {\bibfnamefont {G.}~\bibnamefont {Wolfowicz}},
  \bibinfo {author} {\bibfnamefont {C.~P.}\ \bibnamefont {Anderson}}, \bibinfo
  {author} {\bibfnamefont {A.}~\bibnamefont {Bourassa}}, \bibinfo {author}
  {\bibfnamefont {H.}~\bibnamefont {Ma}}, \bibinfo {author} {\bibfnamefont
  {M.}~\bibnamefont {Ye}}, \bibinfo {author} {\bibfnamefont {G.}~\bibnamefont
  {Koolstra}}, \bibinfo {author} {\bibfnamefont {K.~J.}\ \bibnamefont
  {Satzinger}}, \bibinfo {author} {\bibfnamefont {M.~V.}\ \bibnamefont {Holt}},
  \bibinfo {author} {\bibfnamefont {F.~J.}\ \bibnamefont {Heremans}}, \bibinfo
  {author} {\bibfnamefont {A.~N.}\ \bibnamefont {Cleland}}, \bibinfo {author}
  {\bibfnamefont {D.~I.}\ \bibnamefont {Schuster}}, \bibinfo {author}
  {\bibfnamefont {G.}~\bibnamefont {Galli}},\ and\ \bibinfo {author}
  {\bibfnamefont {D.~D.}\ \bibnamefont {Awschalom}},\ }\href
  {https://doi.org/10.1038/s41567-019-0420-0} {\bibfield  {journal} {\bibinfo
  {journal} {Nat. Phys.}\ }\textbf {\bibinfo {volume} {15}},\ \bibinfo {pages}
  {490} (\bibinfo {year} {2019})}\BibitemShut {NoStop}%
\bibitem [{\citenamefont {Kresse}\ and\ \citenamefont
  {Hafner}(1993)}]{Kresse1993ab}%
  \BibitemOpen
  \bibfield  {author} {\bibinfo {author} {\bibfnamefont {G.}~\bibnamefont
  {Kresse}}\ and\ \bibinfo {author} {\bibfnamefont {J.}~\bibnamefont
  {Hafner}},\ }\href {https://doi.org/10.1103/PhysRevB.47.558} {\bibfield
  {journal} {\bibinfo  {journal} {Phys. Rev. B}\ }\textbf {\bibinfo {volume}
  {47}},\ \bibinfo {pages} {558} (\bibinfo {year} {1993})}\BibitemShut
  {NoStop}%
\bibitem [{\citenamefont {Kresse}\ and\ \citenamefont
  {Furthm\"uller}(1996)}]{Kresse1996efficient}%
  \BibitemOpen
  \bibfield  {author} {\bibinfo {author} {\bibfnamefont {G.}~\bibnamefont
  {Kresse}}\ and\ \bibinfo {author} {\bibfnamefont {J.}~\bibnamefont
  {Furthm\"uller}},\ }\href {https://doi.org/10.1103/PhysRevB.54.11169}
  {\bibfield  {journal} {\bibinfo  {journal} {Phys. Rev. B}\ }\textbf {\bibinfo
  {volume} {54}},\ \bibinfo {pages} {11169} (\bibinfo {year}
  {1996})}\BibitemShut {NoStop}%
\bibitem [{\citenamefont {Kuate~Defo}\ \emph
  {et~al.}(2021{\natexlab{b}})\citenamefont {Kuate~Defo}, \citenamefont
  {Zhang}, \citenamefont {Richardson},\ and\ \citenamefont
  {Kaxiras}}]{Kuate2021theor}%
  \BibitemOpen
  \bibfield  {author} {\bibinfo {author} {\bibfnamefont {R.}~\bibnamefont
  {Kuate~Defo}}, \bibinfo {author} {\bibfnamefont {X.}~\bibnamefont {Zhang}},
  \bibinfo {author} {\bibfnamefont {S.~L.}\ \bibnamefont {Richardson}},\ and\
  \bibinfo {author} {\bibfnamefont {E.}~\bibnamefont {Kaxiras}},\ }\href
  {https://doi.org/10.1063/5.0061396} {\bibfield  {journal} {\bibinfo
  {journal} {J. Appl. Phys.}\ }\textbf {\bibinfo {volume} {130}},\ \bibinfo
  {pages} {155102} (\bibinfo {year} {2021}{\natexlab{b}})}\BibitemShut
  {NoStop}%
\bibitem [{\citenamefont {Kresse}\ and\ \citenamefont
  {Joubert}(1999)}]{Kresse1999from}%
  \BibitemOpen
  \bibfield  {author} {\bibinfo {author} {\bibfnamefont {G.}~\bibnamefont
  {Kresse}}\ and\ \bibinfo {author} {\bibfnamefont {D.}~\bibnamefont
  {Joubert}},\ }\href {https://doi.org/10.1103/PhysRevB.59.1758} {\bibfield
  {journal} {\bibinfo  {journal} {Phys. Rev. B}\ }\textbf {\bibinfo {volume}
  {59}},\ \bibinfo {pages} {1758} (\bibinfo {year} {1999})}\BibitemShut
  {NoStop}%
\bibitem [{\citenamefont {Heyd}\ \emph {et~al.}(2003)\citenamefont {Heyd},
  \citenamefont {Scuseria},\ and\ \citenamefont {Ernzerhof}}]{Heyd}%
  \BibitemOpen
  \bibfield  {author} {\bibinfo {author} {\bibfnamefont {J.}~\bibnamefont
  {Heyd}}, \bibinfo {author} {\bibfnamefont {G.~E.}\ \bibnamefont {Scuseria}},\
  and\ \bibinfo {author} {\bibfnamefont {M.}~\bibnamefont {Ernzerhof}},\ }\href
  {https://doi.org/10.1063/1.1564060} {\bibfield  {journal} {\bibinfo
  {journal} {The Journal of Chemical Physics}\ }\textbf {\bibinfo {volume}
  {118}},\ \bibinfo {pages} {8207} (\bibinfo {year} {2003})}\BibitemShut
  {NoStop}%
\bibitem [{\citenamefont {Krukau}\ \emph {et~al.}(2006)\citenamefont {Krukau},
  \citenamefont {Vydrov}, \citenamefont {Izmaylov},\ and\ \citenamefont
  {Scuseria}}]{Krukau}%
  \BibitemOpen
  \bibfield  {author} {\bibinfo {author} {\bibfnamefont {A.~V.}\ \bibnamefont
  {Krukau}}, \bibinfo {author} {\bibfnamefont {O.~A.}\ \bibnamefont {Vydrov}},
  \bibinfo {author} {\bibfnamefont {A.~F.}\ \bibnamefont {Izmaylov}},\ and\
  \bibinfo {author} {\bibfnamefont {G.~E.}\ \bibnamefont {Scuseria}},\ }\href
  {https://doi.org/10.1063/1.2404663} {\bibfield  {journal} {\bibinfo
  {journal} {The Journal of Chemical Physics}\ }\textbf {\bibinfo {volume}
  {125}},\ \bibinfo {pages} {224106} (\bibinfo {year} {2006})}\BibitemShut
  {NoStop}%
\bibitem [{\citenamefont {Kuate~Defo}\ \emph {et~al.}(2023)\citenamefont
  {Kuate~Defo}, \citenamefont {Rodriguez}, \citenamefont {Kaxiras},\ and\
  \citenamefont {Richardson}}]{Kuate2023theor}%
  \BibitemOpen
  \bibfield  {author} {\bibinfo {author} {\bibfnamefont {R.}~\bibnamefont
  {Kuate~Defo}}, \bibinfo {author} {\bibfnamefont {A.~W.}\ \bibnamefont
  {Rodriguez}}, \bibinfo {author} {\bibfnamefont {E.}~\bibnamefont {Kaxiras}},\
  and\ \bibinfo {author} {\bibfnamefont {S.~L.}\ \bibnamefont {Richardson}},\
  }\href {https://doi.org/10.1103/PhysRevB.107.125305} {\bibfield  {journal}
  {\bibinfo  {journal} {Phys. Rev. B}\ }\textbf {\bibinfo {volume} {107}},\
  \bibinfo {pages} {125305} (\bibinfo {year} {2023})}\BibitemShut {NoStop}%
\bibitem [{\citenamefont {Kaxiras}(2003)}]{Kaxiras2003atomic}%
  \BibitemOpen
  \bibfield  {author} {\bibinfo {author} {\bibfnamefont {E.}~\bibnamefont
  {Kaxiras}},\ }\href {https://doi.org/10.1017/CBO9780511755545} {\emph
  {\bibinfo {title} {Atomic and Electronic Structure of Solids}}}\ (\bibinfo
  {publisher} {Cambridge University Press},\ \bibinfo {year}
  {2003})\BibitemShut {NoStop}%
\bibitem [{\citenamefont {Zhang}\ and\ \citenamefont
  {Northrup}(1991)}]{zhang1991chemical}%
  \BibitemOpen
  \bibfield  {author} {\bibinfo {author} {\bibfnamefont {S.~B.}\ \bibnamefont
  {Zhang}}\ and\ \bibinfo {author} {\bibfnamefont {J.~E.}\ \bibnamefont
  {Northrup}},\ }\href {https://doi.org/10.1103/PhysRevLett.67.2339} {\bibfield
   {journal} {\bibinfo  {journal} {Phys. Rev. Lett.}\ }\textbf {\bibinfo
  {volume} {67}},\ \bibinfo {pages} {2339} (\bibinfo {year}
  {1991})}\BibitemShut {NoStop}%
\bibitem [{\citenamefont {Freysoldt}\ \emph {et~al.}(2014)\citenamefont
  {Freysoldt}, \citenamefont {Grabowski}, \citenamefont {Hickel}, \citenamefont
  {Neugebauer}, \citenamefont {Kresse}, \citenamefont {Janotti},\ and\
  \citenamefont {Van~de Walle}}]{Freysoldt2014first}%
  \BibitemOpen
  \bibfield  {author} {\bibinfo {author} {\bibfnamefont {C.}~\bibnamefont
  {Freysoldt}}, \bibinfo {author} {\bibfnamefont {B.}~\bibnamefont
  {Grabowski}}, \bibinfo {author} {\bibfnamefont {T.}~\bibnamefont {Hickel}},
  \bibinfo {author} {\bibfnamefont {J.}~\bibnamefont {Neugebauer}}, \bibinfo
  {author} {\bibfnamefont {G.}~\bibnamefont {Kresse}}, \bibinfo {author}
  {\bibfnamefont {A.}~\bibnamefont {Janotti}},\ and\ \bibinfo {author}
  {\bibfnamefont {C.~G.}\ \bibnamefont {Van~de Walle}},\ }\href
  {https://doi.org/10.1103/RevModPhys.86.253} {\bibfield  {journal} {\bibinfo
  {journal} {Rev. Mod. Phys.}\ }\textbf {\bibinfo {volume} {86}},\ \bibinfo
  {pages} {253} (\bibinfo {year} {2014})}\BibitemShut {NoStop}%
\bibitem [{\citenamefont {Kuate~Defo}\ \emph {et~al.}(2019)\citenamefont
  {Kuate~Defo}, \citenamefont {Kaxiras},\ and\ \citenamefont
  {Richardson}}]{Kuate2019how}%
  \BibitemOpen
  \bibfield  {author} {\bibinfo {author} {\bibfnamefont {R.}~\bibnamefont
  {Kuate~Defo}}, \bibinfo {author} {\bibfnamefont {E.}~\bibnamefont
  {Kaxiras}},\ and\ \bibinfo {author} {\bibfnamefont {S.~L.}\ \bibnamefont
  {Richardson}},\ }\href {https://doi.org/10.1063/1.5123227} {\bibfield
  {journal} {\bibinfo  {journal} {J. Appl. Phys.}\ }\textbf {\bibinfo {volume}
  {126}},\ \bibinfo {pages} {195103} (\bibinfo {year} {2019})}\BibitemShut
  {NoStop}%
\bibitem [{\citenamefont {Kuate~Defo}\ \emph
  {et~al.}(2021{\natexlab{c}})\citenamefont {Kuate~Defo}, \citenamefont
  {Nguyen}, \citenamefont {Ku},\ and\ \citenamefont
  {Rhone}}]{Kuate2021methods}%
  \BibitemOpen
  \bibfield  {author} {\bibinfo {author} {\bibfnamefont {R.}~\bibnamefont
  {Kuate~Defo}}, \bibinfo {author} {\bibfnamefont {H.}~\bibnamefont {Nguyen}},
  \bibinfo {author} {\bibfnamefont {M.~J.~H.}\ \bibnamefont {Ku}},\ and\
  \bibinfo {author} {\bibfnamefont {T.~D.}\ \bibnamefont {Rhone}},\ }\href
  {https://doi.org/10.1063/5.0048833} {\bibfield  {journal} {\bibinfo
  {journal} {J. Appl. Phys.}\ }\textbf {\bibinfo {volume} {129}},\ \bibinfo
  {pages} {225105} (\bibinfo {year} {2021}{\natexlab{c}})}\BibitemShut
  {NoStop}%
\bibitem [{\citenamefont {Zunger}\ and\ \citenamefont
  {Malyi}(2021)}]{zunger2021under}%
  \BibitemOpen
  \bibfield  {author} {\bibinfo {author} {\bibfnamefont {A.}~\bibnamefont
  {Zunger}}\ and\ \bibinfo {author} {\bibfnamefont {O.~I.}\ \bibnamefont
  {Malyi}},\ }\href {https://doi.org/10.1021/acs.chemrev.0c00608} {\bibfield
  {journal} {\bibinfo  {journal} {Chemical Reviews}\ }\textbf {\bibinfo
  {volume} {121}},\ \bibinfo {pages} {3031} (\bibinfo {year}
  {2021})}\BibitemShut {NoStop}%
\bibitem [{\citenamefont {Yang}\ \emph {et~al.}(2015)\citenamefont {Yang},
  \citenamefont {Yin}, \citenamefont {Park},\ and\ \citenamefont
  {Wei}}]{Yang2015self}%
  \BibitemOpen
  \bibfield  {author} {\bibinfo {author} {\bibfnamefont {J.-H.}\ \bibnamefont
  {Yang}}, \bibinfo {author} {\bibfnamefont {W.-J.}\ \bibnamefont {Yin}},
  \bibinfo {author} {\bibfnamefont {J.-S.}\ \bibnamefont {Park}},\ and\
  \bibinfo {author} {\bibfnamefont {S.-H.}\ \bibnamefont {Wei}},\ }\href
  {https://doi.org/10.1038/srep16977} {\bibfield  {journal} {\bibinfo
  {journal} {Sci. Rep.}\ }\textbf {\bibinfo {volume} {5}},\ \bibinfo {pages}
  {16977 } (\bibinfo {year} {2015})}\BibitemShut {NoStop}%
\bibitem [{\citenamefont {Ashcroft}\ and\ \citenamefont
  {Mermin}(1976)}]{Ashcroft1976solid}%
  \BibitemOpen
  \bibfield  {author} {\bibinfo {author} {\bibfnamefont {N.~W.}\ \bibnamefont
  {Ashcroft}}\ and\ \bibinfo {author} {\bibfnamefont {N.~D.}\ \bibnamefont
  {Mermin}},\ }\href {https://books.google.com/books?id=1C9HAQAAIAAJ} {\emph
  {\bibinfo {title} {Solid State Physics}}},\ HRW international editions\
  (\bibinfo  {publisher} {Holt, Rinehart and Winston},\ \bibinfo {year}
  {1976})\BibitemShut {NoStop}%
\bibitem [{\citenamefont {Sque}\ \emph {et~al.}(2006)\citenamefont {Sque},
  \citenamefont {Jones},\ and\ \citenamefont {Briddon}}]{Sque2006structure}%
  \BibitemOpen
  \bibfield  {author} {\bibinfo {author} {\bibfnamefont {S.~J.}\ \bibnamefont
  {Sque}}, \bibinfo {author} {\bibfnamefont {R.}~\bibnamefont {Jones}},\ and\
  \bibinfo {author} {\bibfnamefont {P.~R.}\ \bibnamefont {Briddon}},\ }\href
  {https://doi.org/10.1103/PhysRevB.73.085313} {\bibfield  {journal} {\bibinfo
  {journal} {Phys. Rev. B}\ }\textbf {\bibinfo {volume} {73}},\ \bibinfo
  {pages} {085313} (\bibinfo {year} {2006})}\BibitemShut {NoStop}%
\bibitem [{\citenamefont {Broadway}\ \emph {et~al.}(2018)\citenamefont
  {Broadway}, \citenamefont {Dontschuk}, \citenamefont {Tsai}, \citenamefont
  {Lillie}, \citenamefont {Lew}, \citenamefont {McCallum}, \citenamefont
  {Johnson}, \citenamefont {Doherty}, \citenamefont {Stacey}, \citenamefont
  {Hollenberg},\ and\ \citenamefont {Tetienne}}]{Broadway2018spat}%
  \BibitemOpen
  \bibfield  {author} {\bibinfo {author} {\bibfnamefont {D.~A.}\ \bibnamefont
  {Broadway}}, \bibinfo {author} {\bibfnamefont {N.}~\bibnamefont {Dontschuk}},
  \bibinfo {author} {\bibfnamefont {A.}~\bibnamefont {Tsai}}, \bibinfo {author}
  {\bibfnamefont {S.~E.}\ \bibnamefont {Lillie}}, \bibinfo {author}
  {\bibfnamefont {C.~T.~K.}\ \bibnamefont {Lew}}, \bibinfo {author}
  {\bibfnamefont {J.~C.}\ \bibnamefont {McCallum}}, \bibinfo {author}
  {\bibfnamefont {B.~C.}\ \bibnamefont {Johnson}}, \bibinfo {author}
  {\bibfnamefont {M.~W.}\ \bibnamefont {Doherty}}, \bibinfo {author}
  {\bibfnamefont {A.}~\bibnamefont {Stacey}}, \bibinfo {author} {\bibfnamefont
  {L.~C.~L.}\ \bibnamefont {Hollenberg}},\ and\ \bibinfo {author}
  {\bibfnamefont {J.~P.}\ \bibnamefont {Tetienne}},\ }\href
  {https://doi.org/10.1038/s41928-018-0130-0} {\bibfield  {journal} {\bibinfo
  {journal} {Nature Electronics}\ }\textbf {\bibinfo {volume} {1}},\ \bibinfo
  {pages} {502} (\bibinfo {year} {2018})}\BibitemShut {NoStop}%
\end{thebibliography}%
\end{document}